\documentclass[aps,preprint,superscriptaddress,showpacs,floatfix]{revtex4-1}

\usepackage{colortbl}
\usepackage{graphicx}
\usepackage{dcolumn}
\usepackage{bm}
\usepackage{CJK}
\usepackage{amsmath,amssymb,latexsym}
\usepackage{multirow,array,booktabs,mathrsfs}
\usepackage{setspace}

\begin{document}

\title{Shape transition with temperature of the pear-shaped nuclei in covariant density functional theory}

\author{Wei Zhang}
\affiliation{Henan Key Laboratory of Ion-beam Bioengineering, Zhengzhou University, Zhengzhou 450052, China}
\affiliation{Key Laboratory of Precision Navigation and Technology, National Time Service Center, Chinese Academy of Sciences, Xi＊an 710600, China}
\affiliation{Molecular Foundry, Lawrence Berkeley National Laboratory, Berkeley 94720, USA}
\author{Yi Fei Niu}\email{nyfster@gmail.com}
\affiliation{ ELI-NP, Horia Hulubei National Institute for Physics and Nuclear Engineering,
30 Reactorului Street, RO-077125, Bucharest-Magurele, Romania}

\begin{abstract}

The shape evolutions of the pear-shaped nuclei $^{224}$Ra and even-even $^{144-154}$Ba with temperature are investigated by
the finite-temperature relativistic mean field theory with the treatment of pairing correlations by the BCS approach.
The free energy surfaces as well as the bulk properties including deformations,
pairing gaps, excitation energy, and specific heat for the global minimum are studied.
For $^{224}$Ra, three discontinuities found in the specific heat curve
indicate the pairing transition at temperature 0.4 MeV, and
two shape transitions at temperatures 0.9 and 1.0 MeV, namely
one from quadrupole-octupole deformed to quadrupole deformed,
and the other from quadrupole deformed to spherical.
Furthermore,
the gaps at $N=$136 and $Z=$88 are responsible for stabilizing
the octupole-deformed global minimum at low temperatures.
Similar pairing transition at $T\sim$0.5 MeV  and shape transitions at $T$=0.5-2.2 MeV
are found for even-even $^{144-154}$Ba.
The transition temperatures are roughly proportional to the corresponding deformations at the ground states.

\end{abstract}


\maketitle

\section{Introduction}

The shapes of atomic nuclei as well as the shape evolution and shape phase transition
between the different shapes have been a hot topic for decades.
At medium excitation energies in heavy-ion fusion,
a completely equilibrated system is formed before the compound nucleus decays by particle or $\gamma$ emission~\cite{Egido1993}.
The canonical description of such a system is characterized by the temperature.
When temperature rises, the shape deformations or superfluidity are expected to wash out~\cite{Egido2000}.
Different from the shape evolution or shape phase transition by changing nucleon numbers,
the shape changes may occur at certain critical temperatures in a single nucleus.
Experimental information about nuclear shape changes
can be obtained by means of the giant dipole resonance (GDR) built on excited states~\cite{GDR}.
The theoretical models of GDR in excited nuclei can be found in Refs.~\cite{Ring1984,Niu2009,Chak2016} and references therein.

For the thermal description, the basic theory is developed in Refs.~\cite{Bloch1958,Sauer1976}.
The shape transition at finite-temperature was first studied in Ref.~\cite{Lee1979}.
The finite-temperature Hartree-Fock approximation was developed~\cite{Brack1974,Quentin1978} and
the dependence of nuclear shape transition on the volume was studied by taking $^{24}$Mg as an example~\cite{Yen1994}.
The finite-temperature Hartree-Fock-Bogoliubov theory was formulated~\cite{Goodman1981}
and then applied to the pairing and shape transitions in rare-earth nuclei~\cite{Goodman1986}.
Using the finite-range density dependent Gogny force and a large configuration space
within the framework of the finite-temperature Hartree-Fock-Bogoliubov (FTHFB) theory~\cite{Egido2000},
varied nuclei, including well quadrupole-deformed nuclei, superdeformed nucleus, and octupole deformed nucleus,
gradually collapse to the spherical shape at certain critical temperatures in the range 1.3-2.7 MeV.
Later the statistical shape fluctuation effects on representative nuclei $^{164}$Er, $^{152}$Dy, and $^{192}$Hg
in the quadrupole degree of freedom were
taken into account with the Landau prescription based on the FTHFB theory~\cite{Egido2003}.
It was found that the deformation parameters start to decrease earlier with temperature
as compared with the plain FTHFB prediction, but shape transition signatures are washed out.
The statistical fluctuations can be treated in
the spirit of the Landau theory~\cite{Levit1984,Egido2003}, or
from a more fundamental point of view by using path integral techniques
such as the static path approximation~\cite{Alhassid1984,Rossignoli1994},
the shell model Monte Carlo method~\cite{Lang1993},
the particle number projected BCS method~\cite{Dang1993,Dang2007,Gambacurta2013}, or the
shell-model-like approach~\cite{Liu2015}.
The temperature also affects the effective mass and the neutron skin~\cite{Yuksel2014}.

In the Ra-Th region, the octupole deformation is involved in shape evolution according to
supportive experimental evidence~\cite{Bizzeti2003,Bizzeti2005,Nature2013}.
A very low-lying negative-parity band, soon merging with the positive-parity one for $J >$ 5,
was observed in nuclei $^{224}$Ra and $^{224}$Th~\cite{Bizzeti2003,Bizzeti2005}.
In 2013, based on the measured strong electric odd-multipole transitions connecting
the low-lying parity doublets, $^{224}$Ra was suggested to be a stable pear-shaped nucleus~\cite{Nature2013}.
Since the nuclear Schiff moment (the electric-dipole distribution weighted by radius squared)
and its resulting atomic electric dipole moment are signatures of time-reversal and parity violation,
and they are expected to be amplified in octupole deformed nuclei,
$^{224}$Ra and its neighbor $^{225}$Ra are of great importance for
physics beyond the Standard Model~\cite{Engel2013}.


It is interesting to explore the thermal properties of the octupole-deformed $^{224}$Ra.
In this work, we aim to investigate the shape evolution
when the temperature rises for $^{224}$Ra in the relativistic mean field (RMF) framework.
The RMF theory, which has achieved great success in describing ground-state
properties of both spherical and deformed nuclei all over the nuclear chart \cite{Ring1996,Vretenar2005,Meng2006}, is also applied to study the shape evolution and phase transitions with temperature.
The finite-temperature relativistic Hartree-Bogoliubov theory~\cite{Niu2013} and relativistic Hartree-Fock-Bogoliubov theory~\cite{Long2015} for spherical nuclei are formulated, and used to study the pairing transitions in hot nuclei.
The relativistic Hartree-BCS theory is applied to study the temperature dependence of shapes and
pairing gaps for $^{166,170}$Er and rare-earth nuclei~\cite{Agrawal2000,Agrawal2001}.
A shape phase transition from prolate to spherical shapes is found at temperatures in the range 1.0-2.7 MeV.
Taking into account the unbound nucleon states,
the temperature dependence of the pairing gaps, nuclear deformation, radii, binding energies, and entropy
are studied using the Dirac-Hartree-Bogoliubov (DHB) calculations~\cite{Bonche1984,Lisboa2016}.
It is found that the nuclear deformation disappears at temperatures $T$ = 2.0-4.0 MeV.
When the temperature $T \geqslant 4$ MeV, the effects of the vapor phase that
take into account the unbound nucleon states become important.
Considering that different shape phase transition temperatures are found for varied nuclei
in the covariant density functional framework,
it is timely to discuss the transition temperatures for this typical octupole deformed nucleus $^{224}$Ra
with a novel point-coupling parameter set PC-PK1.

The paper will be organized as follows.
The self-consistent finite-temperature RMF theory with BCS approach for axially deformed nuclei
based on the point-coupling density functional will be briefly presented.
After that, the free energy surface, the quadrupole and octupole deformations, the excitation energy, and
the specific heat as functions of the temperature will be discussed.
Finally the evolution of the single-particle spectra will be shown.


\section{Theoretical framework}

The starting point of the RMF theory is an effective Lagrangian density
with zero-range point-coupling interaction between nucleons:
\begin{eqnarray}\label{Eq:Lagrangian}
 \mathcal{L}&=& \bar\psi(i\gamma_\mu\partial^\mu-m)\psi \nonumber\\
            && -\frac{1}{2}\alpha_S(\bar\psi\psi)(\bar\psi\psi)
               -\frac{1}{2}\alpha_{V}(\bar\psi\gamma_\mu\psi)(\bar\psi\gamma^\mu\psi)
               -\frac{1}{2}\alpha_{TV}(\bar\psi\vec\tau\gamma_\mu\psi)\cdot(\bar\psi\vec\tau\gamma^\mu\psi) \nonumber\\
            && -\frac{1}{3}\beta_S(\bar\psi\psi)^3-\frac{1}{4}\gamma_S(\bar\psi\psi)^4
               -\frac{1}{4}\gamma_V[(\bar\psi\gamma_\mu\psi)(\bar\psi\gamma^\mu\psi)]^2 \nonumber\\
            && -\frac{1}{2}\delta_S\partial_\nu(\bar\psi\psi)\partial^\nu(\bar\psi\psi)
               -\frac{1}{2}\delta_V\partial_\nu(\bar\psi\gamma_\mu\psi)\partial^\nu(\bar\psi\gamma^\mu\psi) \nonumber\\
            && -\frac{1}{2}\delta_{TV}\partial_\nu(\bar\psi\vec\tau\gamma_\mu\psi)\cdot\partial^\nu(\bar\psi\vec\tau\gamma^\mu\psi)\nonumber\\
            && -\frac{1}{4}F^{\mu\nu}F_{\mu\nu}  - e\bar\psi\gamma^\mu\frac{1-\tau_3}{2}\psi A_\mu,
\end{eqnarray}
which includes the free-nucleons term, the four-fermion point-coupling terms,
the higher-order terms which are responsible for the effects of medium dependence,
the gradient terms which are included to simulate the effects of finite range,
and the electromagnetic interaction terms.
The isovector-scalar channel is neglected.
The Dirac spinor field of the nucleon is denoted by $\psi$, and the nucleon mass is $m$.
$\vec\tau$ is the isospin Pauli matrix, and $\Gamma$ generally denotes the 4$\times$4 Dirac matrices
including $\gamma_\mu$, $\sigma_{\mu\nu}$ while
greek indices $\mu$ and $\nu$ run over the Minkowski indices 0, 1, 2, and 3.
$\alpha$, $\beta$, $\gamma$, and $\delta$ with subscripts $S$ (scalar), $V$ (vector), $TV$ (isovector) are coupling constants (adjustable parameters) in which $\alpha$ refers to the four-fermion term, $\beta$ and $\gamma$ respectively to the third- and fourth-order terms, and $\delta$ the derivative couplings.

Following the prescription in Ref.~\cite{Goodman1981} where the BCS limit of finite-temperature
Hartree-Fock Bogoliubov equations is derived, we obtain the finite-temperature RMF + BCS equation.
The finite-temperature Dirac equation for single nucleons reads~\cite{Kr2017}
\begin{equation}\label{Eq:Dirac-PC}
  [\gamma_\mu(i\partial^\mu-V^\mu)-(m+S)]\psi_k=0,
\end{equation}
where $m$ is the nucleon mass.  $\psi_k(\bm{r})$ denotes the Dirac spinor field of a nucleon.
The scalar $S(\bm{r})$ and vector $V^\mu(\bm{r})$ potentials are
\begin{equation}\label{Eq:S}
S(\bm{r})  =\alpha_S\rho_S+\beta_S\rho^2_S+\gamma_S\rho^3_S+\delta_S\triangle\rho_S,
\end{equation}
\begin{eqnarray}\label{Eq:V}
V^\mu (\bm{r}) &=&\alpha_Vj^\mu_V +\gamma_V (j^\mu_V)^3 +\delta_V\triangle j^\mu_V\nonumber\\
               & &+\tau_3\alpha_{TV} \vec{j}^\mu_{TV}+ \tau_3\delta_{TV}\triangle \vec{j}^\mu_{TV}+ e A^\mu
\end{eqnarray}
respectively.
The isoscalar density $\rho_S$, isoscalar current $j^\mu_V$ and isovector current $\vec{j}^\mu_{TV}$ are
represented by,
\begin{eqnarray}
\rho_S (\bm{r})      &=& \sum \limits_{k} \bar\psi_k(\bm{r}) \psi_k(\bm{r}) [v_k^2 (1-2 f_k)+f_k],\label{Eq:dencur1} \\
j^\mu_V (\bm{r})     &=& \sum \limits_{k} \bar\psi_k(\bm{r}) \gamma^\mu \psi_k(\bm{r}) [v_k^2 (1-2 f_k)+f_k], \label{Eq:dencur2}\\
\vec{j}^\mu_{TV} (\bm{r}) &=& \sum \limits_{k} \bar\psi_k(\bm{r}) \vec{\tau} \gamma^\mu \psi_k(\bm{r}) [v_k^2 (1-2 f_k)+f_k].\label{Eq:dencur3}
\end{eqnarray}
where the thermal occupation probability of quasiparticle states $f_k$
is directly related to the temperature $T$ by $f_k=1/(1+e^{E_k/k_BT})$.
$E_k$ is the quasiparticle energy for single-particle (s.p.) state $k$,
$E_k = [(\epsilon_k-\lambda_q)^2 +( \Delta_k)^2]^{\frac{1}{2}}$.
In Eqs.~(\ref{Eq:dencur1})-(\ref{Eq:dencur3}),
the BCS occupation probabilities $v_k^2$ and associated $u_k^2=1-v_k^2$ are obtained by
\begin{eqnarray}\label{eq:occ}
v_k^2   &=&\frac{1}{2} (1- \frac{\epsilon_k-\lambda_q}{E_k}), \\
u_k^2   &=&\frac{1}{2} (1+ \frac{\epsilon_k-\lambda_q}{E_k}).
\end{eqnarray}
$\Delta_k$ is the pairing gap parameter, which satisfies the gap equation at finite temperature.
\begin{equation}
  \Delta_k = - \frac{1}{2} \sum_{k'>0} V^{pp}_{k\bar{k} k' \bar{k}'} \frac{\Delta_{k'}}{ E_{k'}} (1-2f_{k'}).
\end{equation}
The particle number $N_q$ is restricted by $N_q= 2 \sum \limits_{k>0} [v_k^2 (1-2 f_k)+f_k]$.

Here we take the $\delta$ pairing force $V(\bm{r})=V_q\delta(\bm{r})$, where $V_q$ is the pairing strength parameter for neutrons or protons.
A smooth energy-dependent cutoff weight $g_k$ having the form $\{1+\exp[(\epsilon_k-\lambda_q-E_c)/(E_c/10)]\}^{-1}$,
where $E_c$ is the cutoff parameter, is introduced to simulate the effect of finite range
in the evaluation of local pair density.
The cutoff parameter is determined by an approximate condition $\sum \limits_{k} 2 g_k= N_q + 1.65 N^{2/3}_q $ related to the particle number $N_q$, where $q$ can refer to neutron or proton~\cite{Bender2000}.

The internal binding energies $E$ at different axial-symmetric shapes
can be obtained by applying constraints with quadrupole deformation $\beta_2$ and octupole deformation $\beta_3$ together.
In the mean-field level, the binding energy with a given deformation $\beta_2$ can be obtained by minimizing
\begin{equation}\label{eq:def-cons}
    \langle H'\rangle=\langle H\rangle +\frac{1}{2}C(\langle \hat{Q}_2\rangle-\mu_2)^2,
\end{equation}
where $C$ is a spring constant, $\mu_2 =\frac {3AR^2} {4\pi} \beta_2$ is the given quadrupole moment, and $\langle \hat{Q}_2\rangle$ is the expectation value of qudrupole moment operator $\hat{Q}_2=2r^2P_2(\cos\theta)$. The octupole moment constraint can also be applied similarly with $\hat{Q}_3=2r^3P_3(\cos\theta)$ and $\mu_3 =\frac {3AR^3} {4\pi} \beta_3$.

The free energy for the system is $F=E-TS$ where the entropy $S$ is evaluated from
\begin{equation}\label{eq:S}
S=-k_B \sum \limits_{k} [f_k{\rm ln}f_k +(1-f_k){\rm ln}(1-f_k)].
\end{equation}
For convenience, the temperature used is $k_BT$ in units of MeV and
the entropy used is $S/k_B$ and is unitless.
The free energy surface in the ($\beta_2, \beta_3$) plane is obtained by plotting
the free energy $E(\beta_2, \beta_3)-TS(\beta_2, \beta_3)$ on a mesh with equidistant $\beta_2$ and $\beta_3$.

The specific heat is defined by the relation
\begin{equation}
C_{\rm v}=\partial E^*/\partial T,
\end{equation}
where $E^*(T)=E(T)-E(T=0)$ is the internal excitation energy, and
$E(T)$ is the internal binding energy for the global minimum state in the free energy surface at certain temperature $T$.

\section{Results and discussion}

The point-coupling density functional parameter set PC-PK1 is used in our calculation
due to its success in the description of finite nuclei for both the ground state and low-lying excited states~\cite{Zhao2010}.
The pairing correlations are taken into account by the BCS method with a $\delta$ pairing force.
The value of the pairing strength for neutrons (protons) $V_q$ is taken from Ref.~\cite{Zhao2010},
that is, -349.5 (-330.0) MeV fm$^3$.
A set of axial harmonic oscillator basis functions with 20 major shells is used.

\begin{figure}[htb]
\includegraphics[scale=0.3]{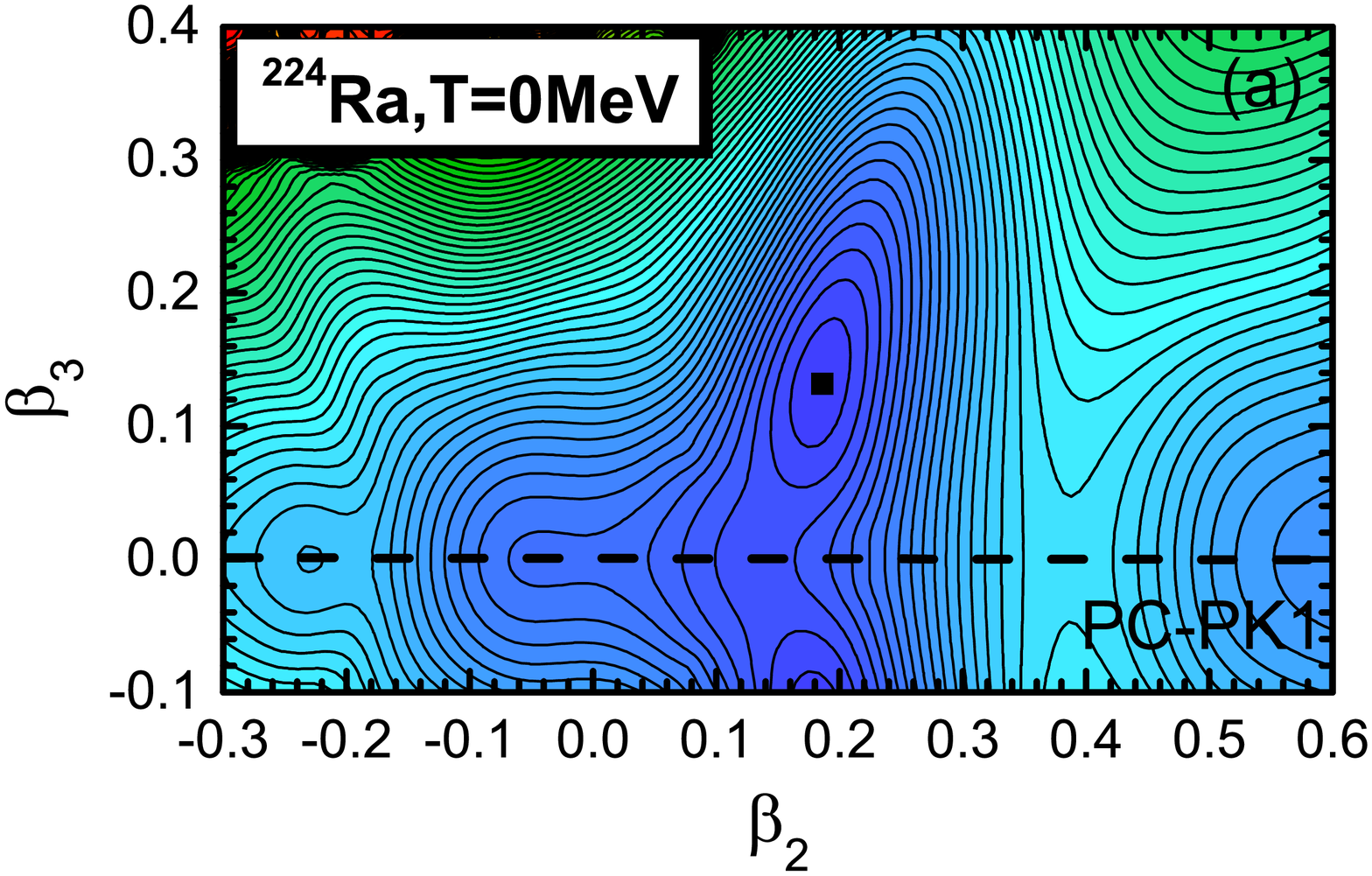} %
\includegraphics[scale=0.3]{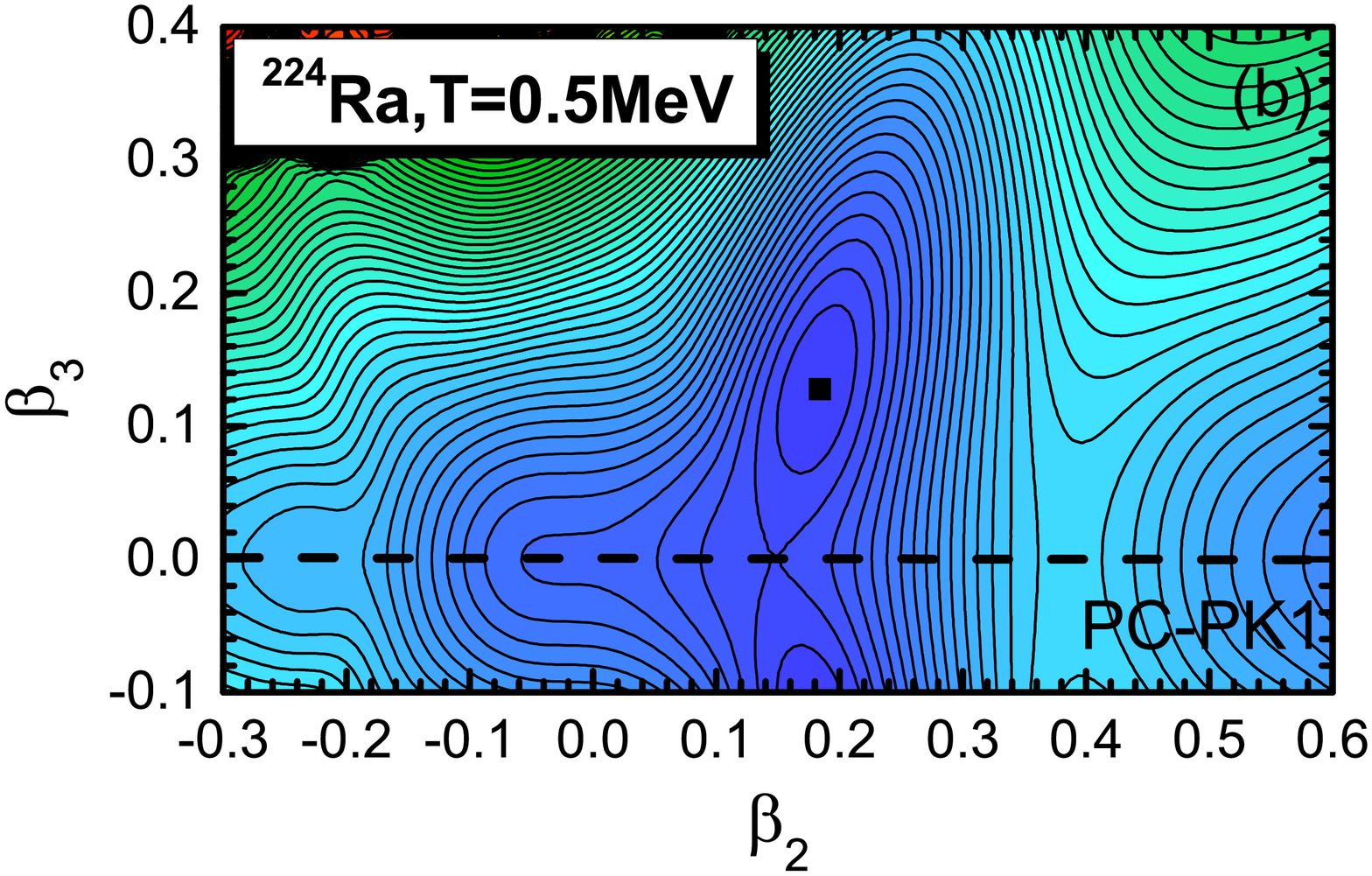} \\
\includegraphics[scale=0.3]{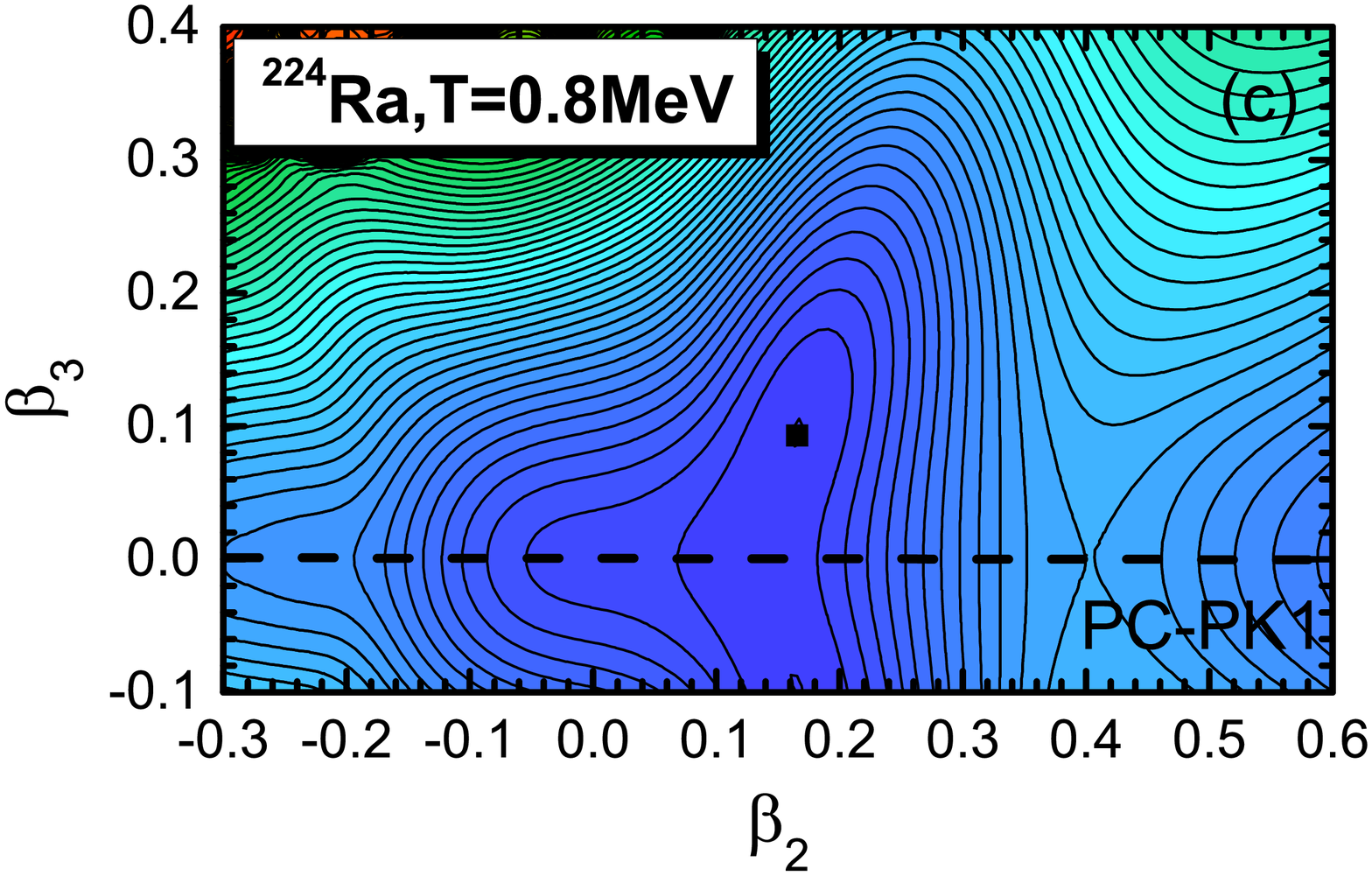} %
\includegraphics[scale=0.3]{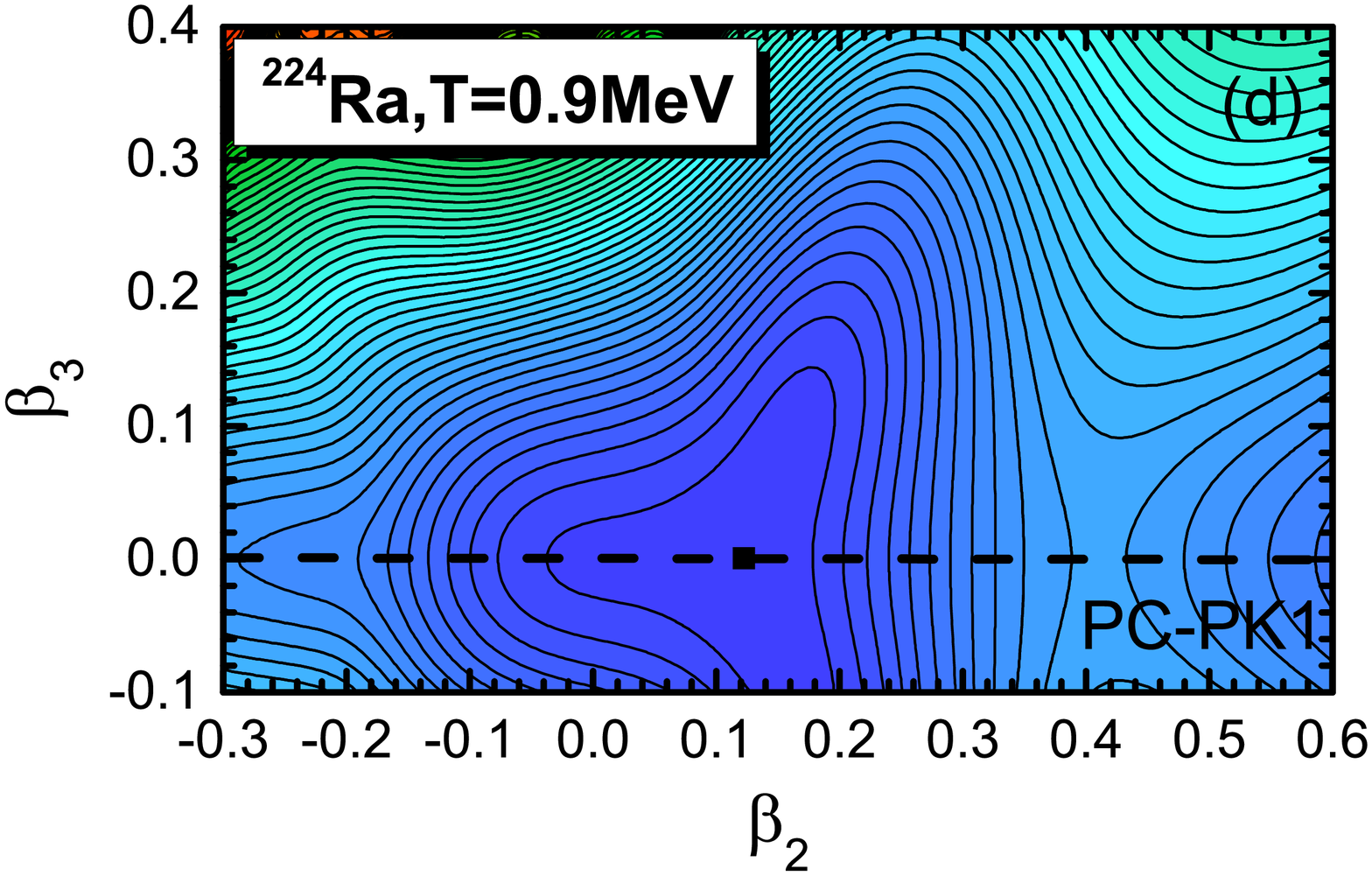} \\
\includegraphics[scale=0.3]{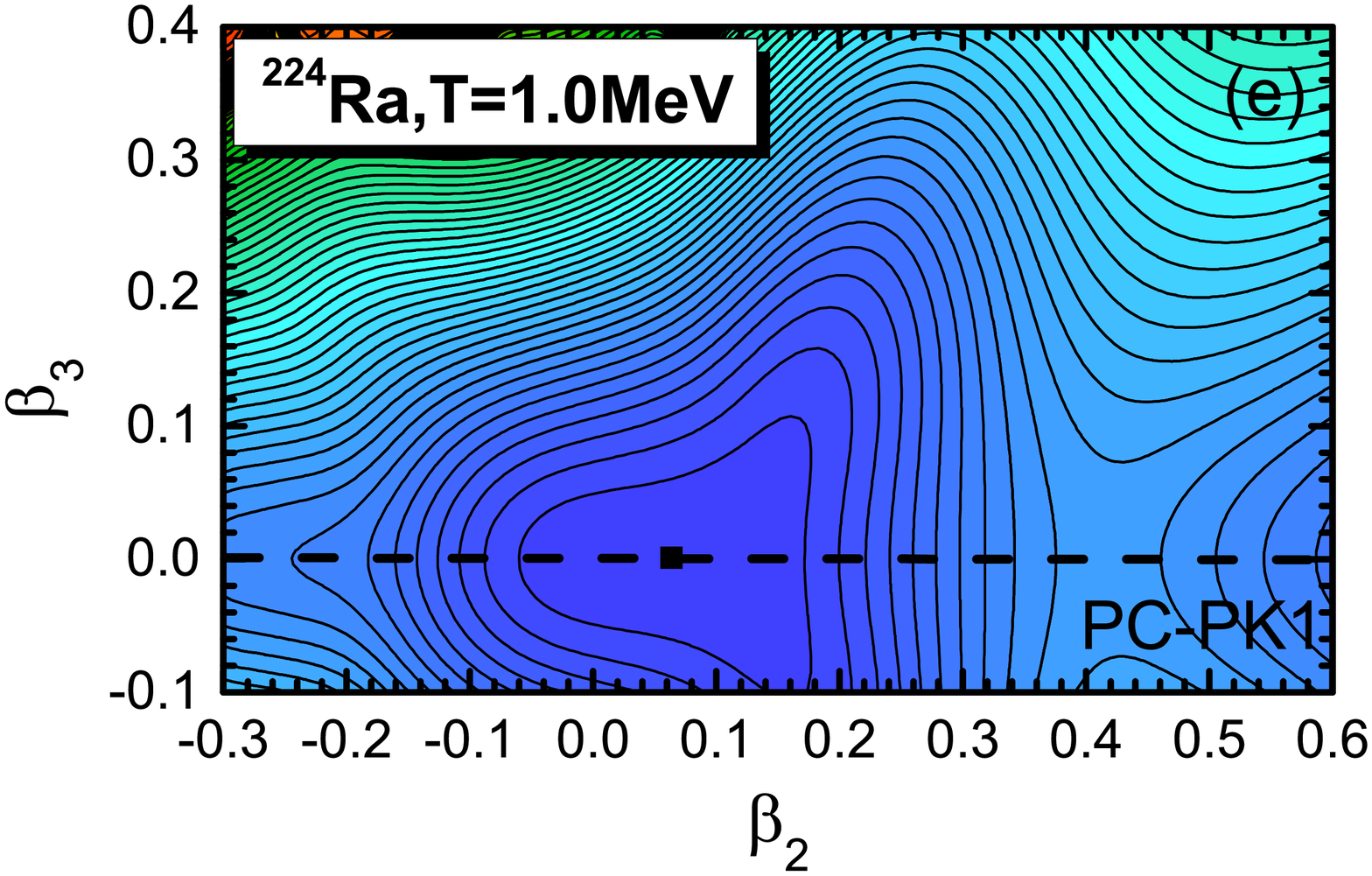} %
\includegraphics[scale=0.3]{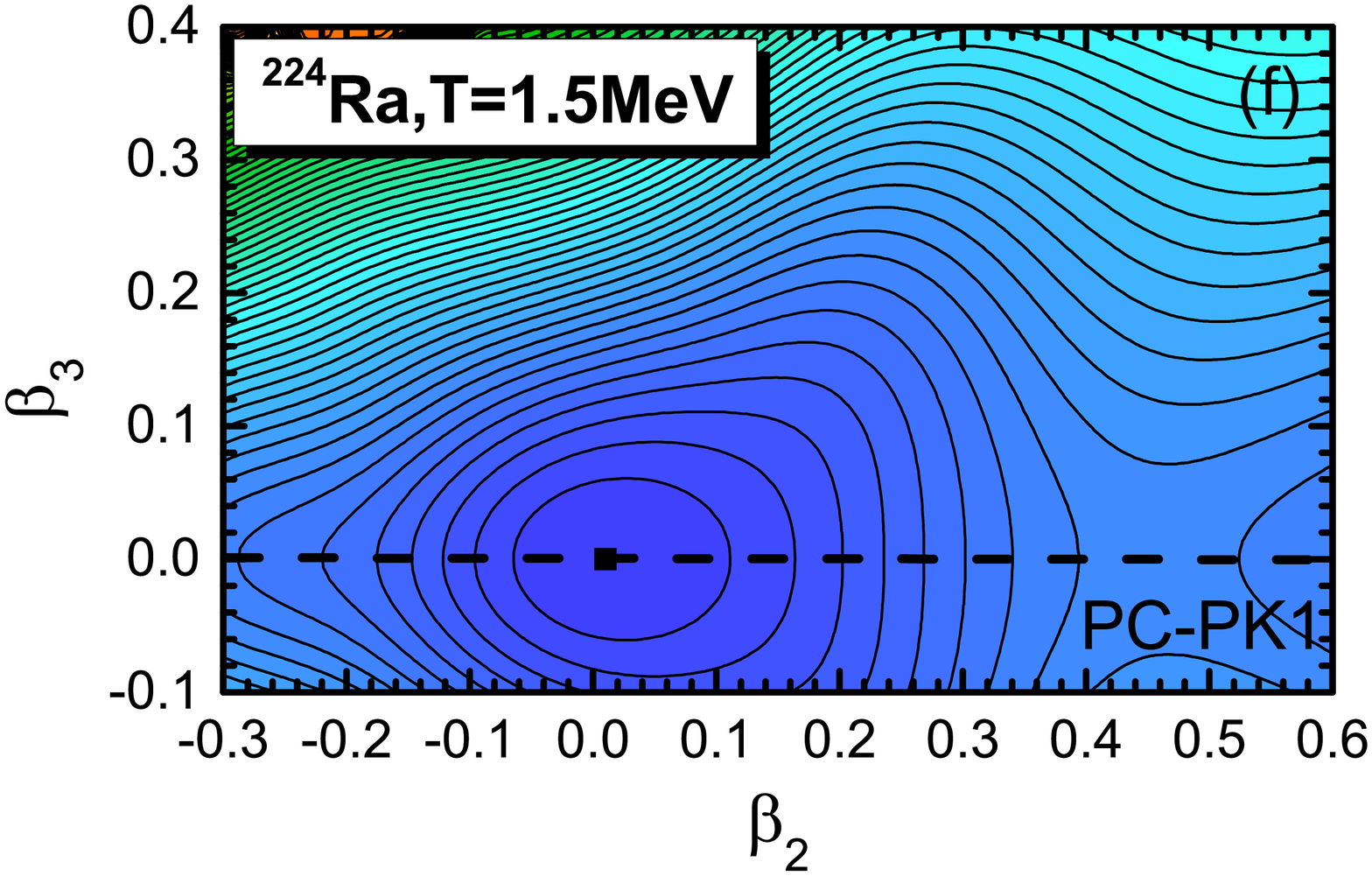} \\
\caption{(Color online) The free energy surfaces in the ($\beta_2$, $\beta_3$) plane at temperatures
T=0 (a), 0.5 (b), 0.8 (c), 0.9 (d), 1.0 (e), and 1.5 (f) MeV for $^{224}$Ra
obtained by the finite-temperature RMF+BCS calculations using the PC-PK1 energy density functional.
The global minima are indicated by the solid squares.
The energy separation between contour lines is 0.5 MeV.
}
\label{pes}
\end{figure}

The free energies in the ($\beta_2$, $\beta_3$) plane at temperatures
$T=$0, 0.5, 0.8, 0.9, 1.0, and 1.5 MeV for $^{224}$Ra are plotted in Fig.~\ref{pes}.
For zero temperature in Fig.~\ref{pes}(a),
the free energy for the global minimum at zero temperature, which equals
the internal binding energy for the ground state,
is close to the experimental binding energy for $^{224}$Ra within 0.1\%.
The corresponding deformations $\beta_2$=0.184, $\beta_3$=0.133 are slightly bigger than
the experimental data $\beta_2$=0.154, $\beta_3$=0.097 in Ref.~\cite{Nature2013}.
Such a result is also consistent with other theoretical calculations, e.g.,
$\beta_2$=0.179, $\beta_3$=0.125 for parameter set PC-PK1 in Ref.~\cite{Yao2015} and
$\beta_2$=0.177, $\beta_3$=0.125 for parameter set DD-PC1,
$\beta_2$=0.178, $\beta_3$=0.124 for parameter set NL3* in Ref.~\cite{Ring2016}.
The saddle point is purely quadrupole deformed, about 1.1 MeV higher than the global minimum.
When the temperature rises up to 0.5 MeV, the deformations of the global minimum changes little.
In the temperature range $0.5 \leqslant T \leqslant 1.0 $MeV, the energy surfaces change dramatically.
The energy difference between the saddle point and global minimum gradually decreases, and finally drops to zero.
At $T \sim$ 0.9 MeV, the saddle point and global minimum merge together at $\beta_2$=0.12, and $\beta_3$=0,
and the shape phase transition from octupole deformed to quadrupole deformed occurs.
For $0.8 \leqslant T \leqslant 1.0 $ MeV, a relatively soft area near the global minimum is composed;
e.g., at $T=0.9$MeV the states with $-0.04 \leqslant \beta_2 \leqslant 0.18$, $\beta_3=0$, or
$\beta_2=0.19$, $\beta_3=0.14$ are no more than 0.5 MeV higher than the global minimum.
At $T \sim$ 1.0 MeV, the global minimum moves to a spheroidal shape at $\beta_2$=0.02, and $\beta_3$=0,
and the nucleus has another shape phase transition from quadrupole deformed to spherical shape.
When the temperature continues rising above $1.5$ MeV, the energy surfaces evolves little
with a single minimum near spherical state.

\begin{figure}[htb]
\includegraphics[scale=0.6]{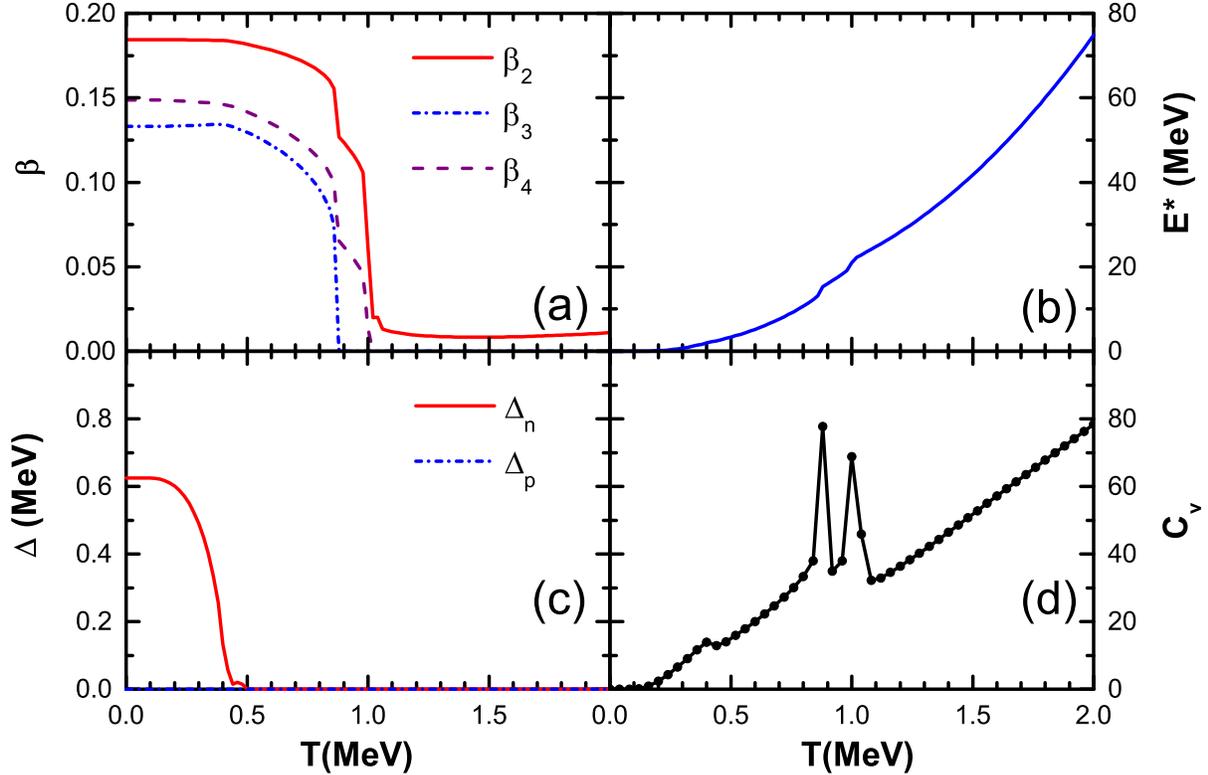}
\caption{
(Color online) The global minima deformations $\beta_2$, $\beta_3$, $\beta_4$ (a), excitation energies $E^*$ (in MeV) (b), pairing gaps $\Delta_n$, $\Delta_p$ (in MeV) (c), and the specific heat $C_{\rm v}$ (d) as functions of temperature (in MeV) for $^{224}$Ra, obtained by the finite-temperature RMF+BCS calculations using the PC-PK1 energy density functional.
}
\label{basic}
\end{figure}

\begin{figure}[htb]
\includegraphics[scale=0.4]{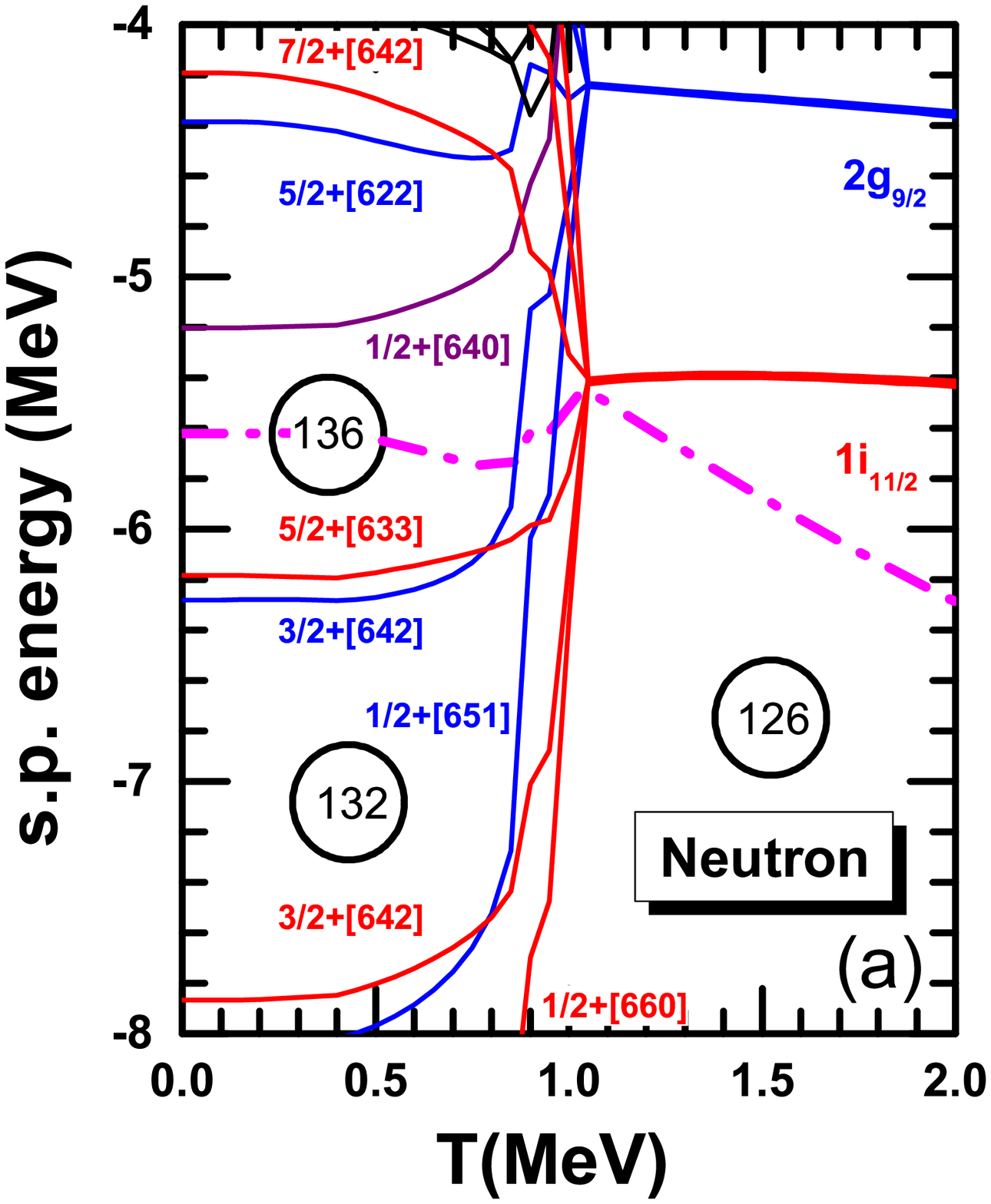}%
\includegraphics[scale=0.4]{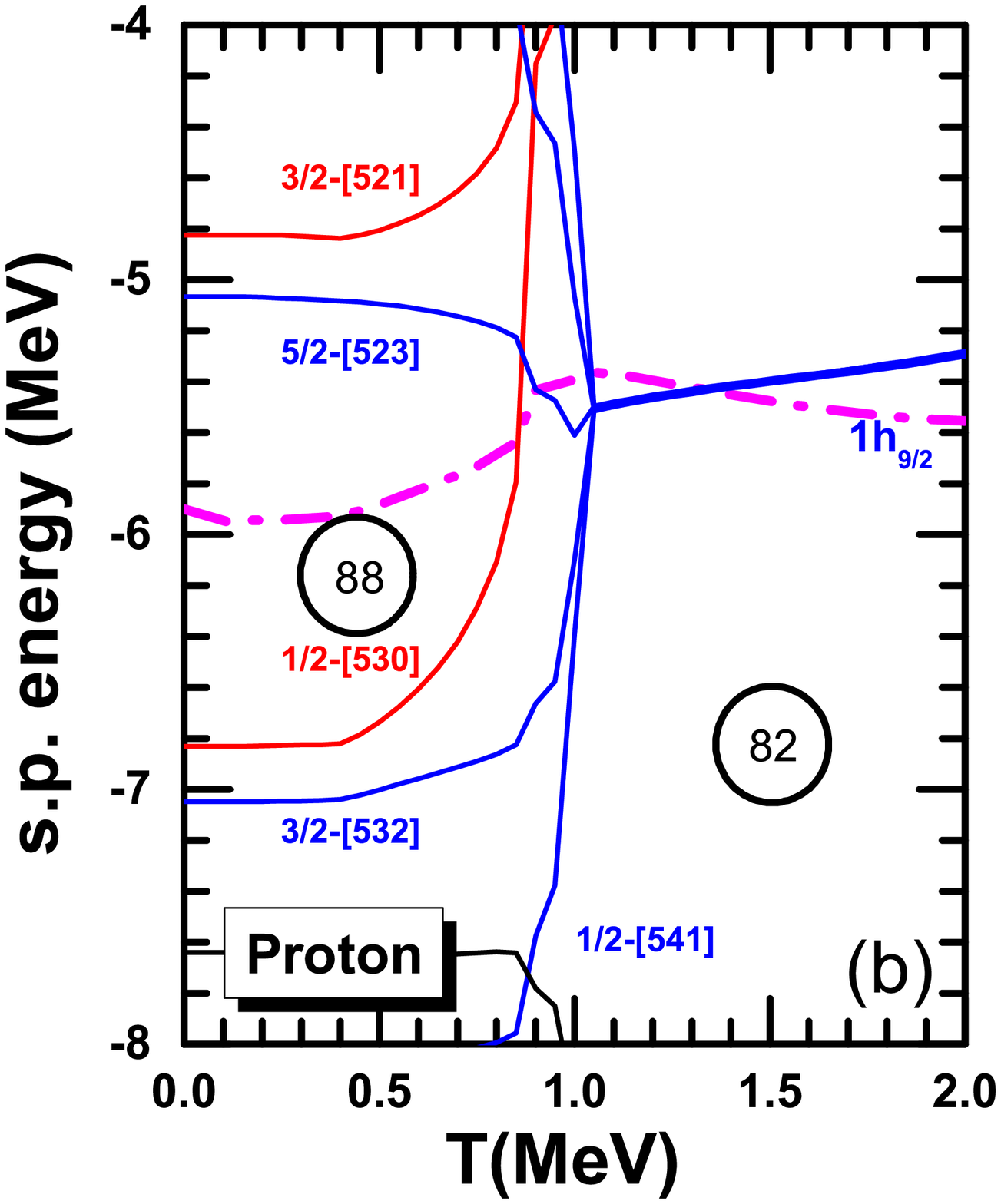}\\
\caption{
(Color online) Neutron (a) and proton (b) single-particle levels as a function of temperature (in MeV) for the nucleus $^{224}$Ra,  obtained by constrained RMF+BCS calculations using the PC-PK1 energy density functional.
The dash-dot lines denote the corresponding Fermi surfaces.
The levels near the Fermi surface are labeled by
Nilsson notations $\Omega$$\pi$[$N$$n_z$$m_l$] of the first leading component.
}
\label{sp}
\end{figure}

To see the properties of the nuclei as functions of temperatures in more detail,
the evolution of deformations, pairing gaps, the excitation energy as well as the specific heat
for $^{224}$Ra using the PC-PK1 energy density functional are shown in Fig.~\ref{basic}.
In Fig.~\ref{basic}(a), the global minimum deformations are almost constant for $T\leqslant 0.5$MeV.
At temperatures 0.9 and 1.0 MeV, the quadrupole and octupole deformations quickly drop to zero respectively.
The hexadecapole deformation behaves similar to the quadrupole deformation.
So the nucleus first goes through a shape transition from octupole to quadrupole, and then another shape transition from quadrupole and hexadecapole deformed to spherical.
The vanishing of octupole deformation also interplays with quadrupole and hexadecapole deformations, which causes sudden decreases in the quadrupole and hexadecaple deformations.
In Fig.~\ref{basic}(c),
the neutron pairing gap at zero temperature is $\Delta_n(0)=0.63$ MeV,
and it decreases to zero at a critical temperature $T_c=0.40$MeV, which shows a pairing phase transition.
The critical temperature for the pairing phase transition in the case of deformed nuclei basically follows
the rule $T_c = 0.6\Delta_n(0)$, which was discovered for spherical nuclei~\cite{Niu2013}.
Since the global minimum deformation for $T\leqslant T_c$ changes little, as seen in Fig.~\ref{basic}(a),
the single-particle levels, which are sensitive to the deformation but insensitive to the temperature,
are barely altered for $T\leqslant T_c$.
It is speculated that the rule $T_c = 0.6\Delta(0)$ is kept for a fixed single-particle level structure,
no matter whether the nucleus prefers a spherical, oblate, or prolate minimum.
Such a rule may not hold for a gradually changing minimum or oscillating minimum.
In Fig.~\ref{basic}(c), the proton pairing gap is zero.
Such a proton pairing collapse is caused by a larger gap near the Fermi surface and weaker pairing strength
compared with the neutron counterpart.
At zero temperature, the proton shell gap near the Fermi surface is about 1.7 MeV
while the corresponding neutron shell gap is 1 MeV (cf. Fig.~\ref{sp}).
Moreover, for PC-PK1 parameter set, the proton pairing strength -330 MeV fm$^3$
is about 5.6\% weaker than the corresponding neutron pairing strength -349.5 MeV fm$^3$.
Compared with Ref.~\cite{Egido2000}, where the proton pairing energy is
a nonzero value -2 MeV,
our pairing strength is in general smaller, which we could see by comparing the neutron pairing energy.
In Ref.~\cite{Egido2000} the neutron pairing energy at zero temperature is -5 MeV, while it is -3.2 MeV in our calculation.

For the relative excitation energy $E^*$ in Fig.~\ref{basic}(b),
one inappreciable kink at low temperatures and two noticeable kinks at medium temperatures can be found.
Its derivative, namely the specific heat is plotted in Fig.~\ref{basic}(d).
Three discontinuities in Fig.~\ref{basic}(d) correspond to three kinks in Fig.~\ref{basic}(b).
It is well-known that the appearance of discontinuities in this quantity is customarily
interpreted as a signature of transitions.
The discontinuity at $T_c=0.4$ MeV indicates the transition from the superfluid to the normal phase.
Two peaks at 0.9 and 1.0 MeV indicate two shape transitions, namely
one from quadrupole-octupole deformed to quadrupole deformed (octupole transition for short),
and the other from quadrupole deformed to spherical (quadrupole transition for short).
However, in experiments the specific heat usually exhibits a more smooth behavior
than compared to the sharp discontinuity obtained here.
This is attributed to the finite size of the nucleus and, therefore,
a realistic description of statistical and quantal fluctuations.
The two adjacent shape transitions would be smeared into one.
In the finite-temperature HFB theory~\cite{Egido2000},
$^{224}$Ra experiences similar transitions:
first a pairing transition at $T=0.5$ MeV,
and after that, the octupole deformation continuously decreases, and finally becomes zero at $T=1.3$ MeV
together with abrupt changes in quadrupole and hexadecapole deformations~\cite{Egido2000}.
The signature of such transition at $T=1.3$ MeV can also be found in the corresponding specific heat.
It is found that for $^{224}$Ra, the proton single-particle gap of $Z$=88
at the deformed equilibrium in the RMF calculation is relatively smaller than
that of the corresponding Gogny D1S calculation~\cite{Robledo2010}.
With a smaller shell gap, the shape transition can be achieved at a lower temperature, which is 1.0 MeV in the RMF calculation compared to 1.3 MeV in the Gogny calculation~\cite{Egido2000}.
Such a smaller shell gap near $Z$=88 is related to the artificial shell closure $Z$=92 (above $1h_{9/2}$) in the spherical state, which is commonly found in RMF calculations, and can be cured by the presence of the degree of freedom associated with the Lorentz tensor $\rho$-field~\cite{Long2007,Meng2016}.

In Fig.~\ref{sp}, we plot the single-particle (s.p.) levels of neutrons and protons at the global minimum as a function of temperature for $^{224}$Ra.
Normally the octupole interaction couples the pairs of orbitals
with $\Delta N=1, \Delta l=3$, and $\Delta j=3$ near the Fermi surface.
The regions of nuclei with strong octupole
correlations correspond to either the proton or neutron numbers being
close to 34 ($1g_{9/2} \leftrightarrow 2p_{3/2}$ coupling), 56 ($1h_{11/2} \leftrightarrow 2d_{5/2}$ coupling),
88 ($1i_{13/2} \leftrightarrow 2f_{7/2}$ coupling), and 134 ($1j_{15/2} \leftrightarrow 2g_{9/2}$ coupling)~\cite{Butler1996}.
A gap at $N=$136, together with another gap at $N=$132 for $T <$ 1 MeV and a big gap at $N=$126 for $T > $1 MeV can be
found in Fig.~\ref{sp}(a) while
a gap at $Z=$88 for $T <$ 1 MeV and a big gap at $Z=$82 for $T > 1$ MeV can be found in Fig.~\ref{sp}(b).
Since the s.p. levels are sensitive to slight deformations,
for the spheroidal global minimum when $T > 1$ MeV,
the s.p. levels constituting $\nu 2g_{9/2}$, $\nu 1i_{11/2}$, and $\pi 1h_{9/2}$ do not degenerate rigorously.
The gaps at $N=$136 and $Z=$88 are responsible to the octupole-deformed global minimum at low temperatures.
With temperature increasing, more and more nucleons are thermally excited to s.p. levels above the shell gaps,
so the shell effect becomes weaker and disappears eventually.
As a result, the shape of the nucleus becomes spherical at high temperature.

\begin{figure}[htb]
\includegraphics[scale=0.2]{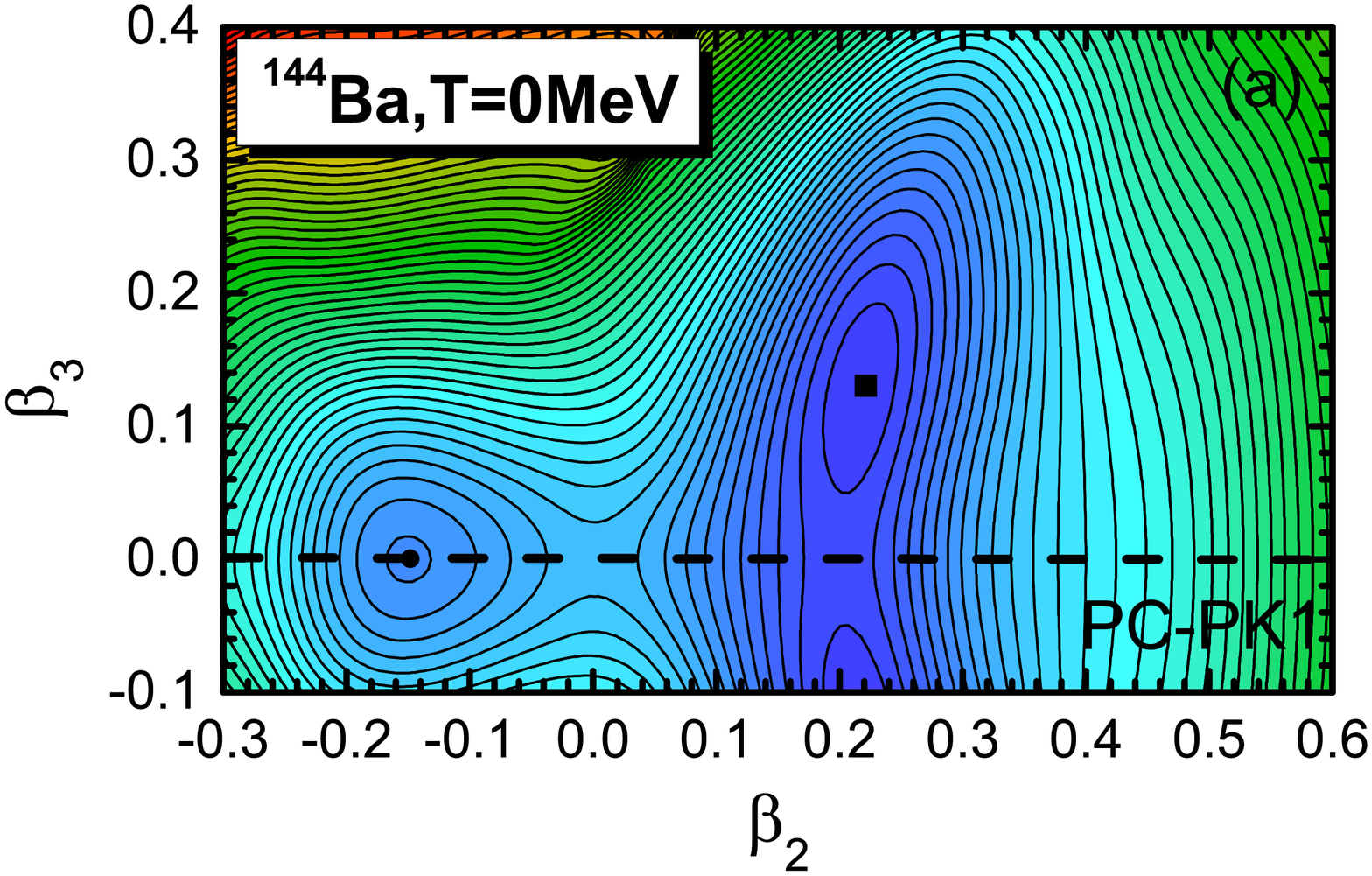} %
\includegraphics[scale=0.2]{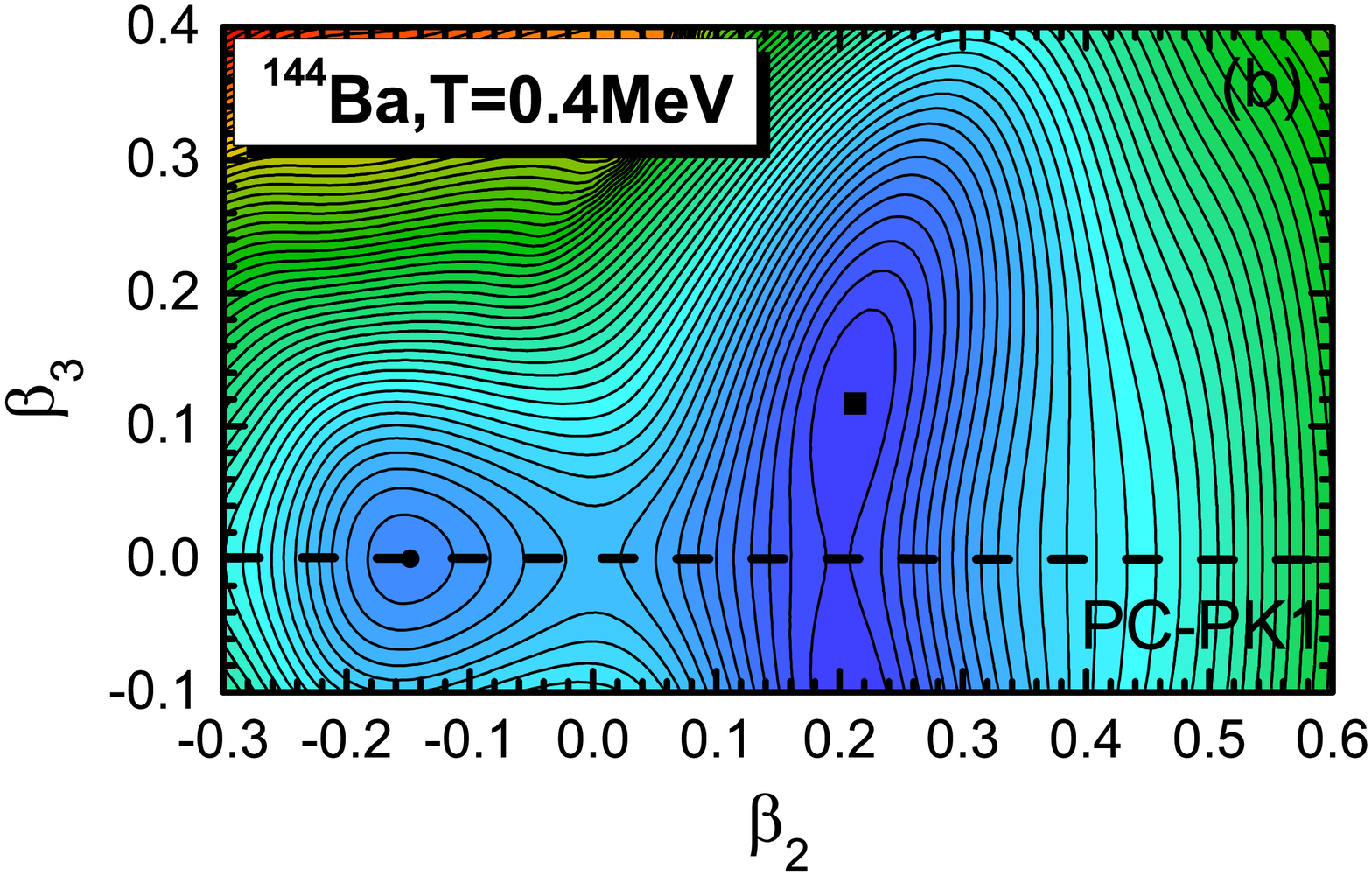} %
\includegraphics[scale=0.2]{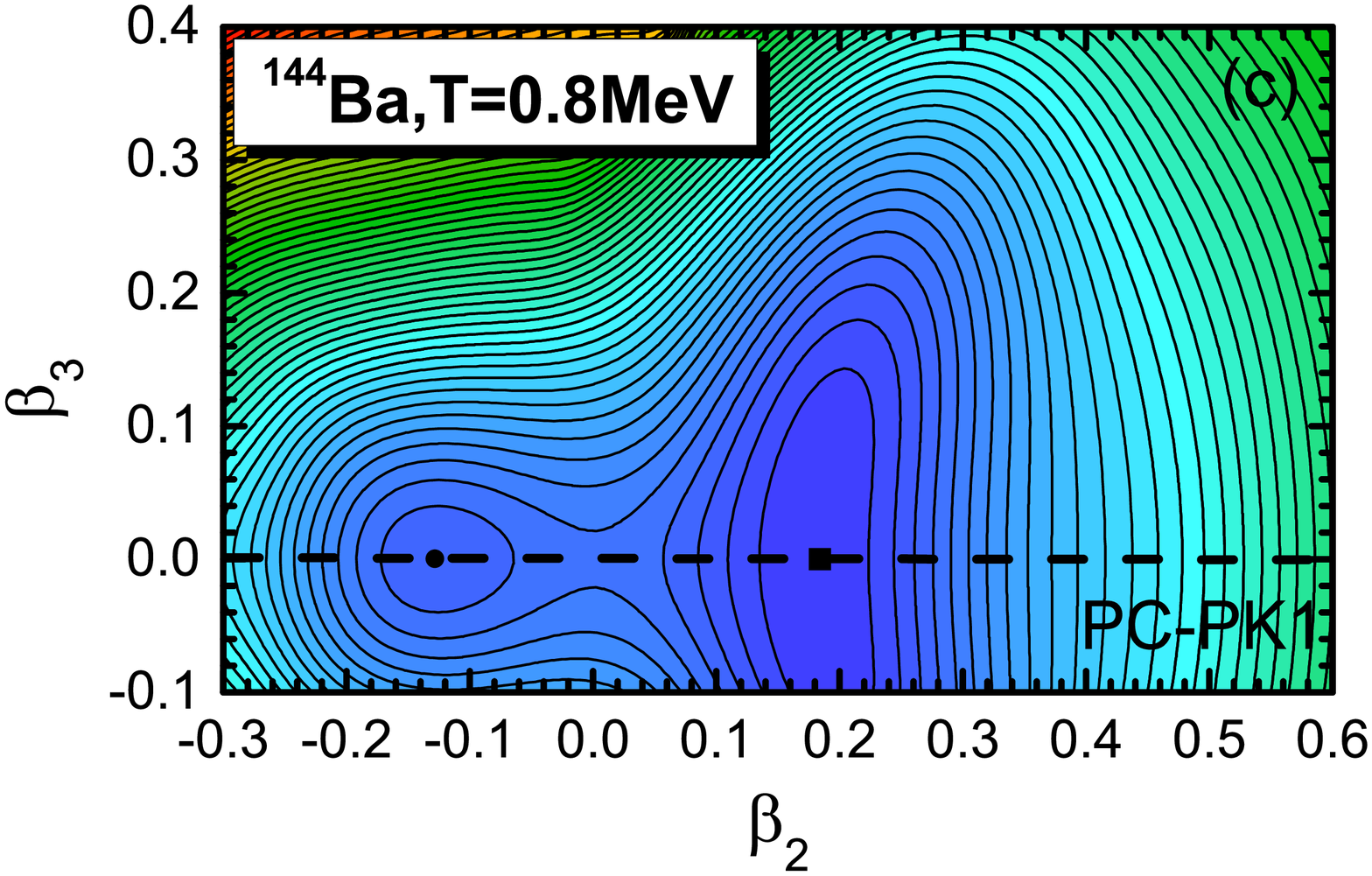} \\
\includegraphics[scale=0.2]{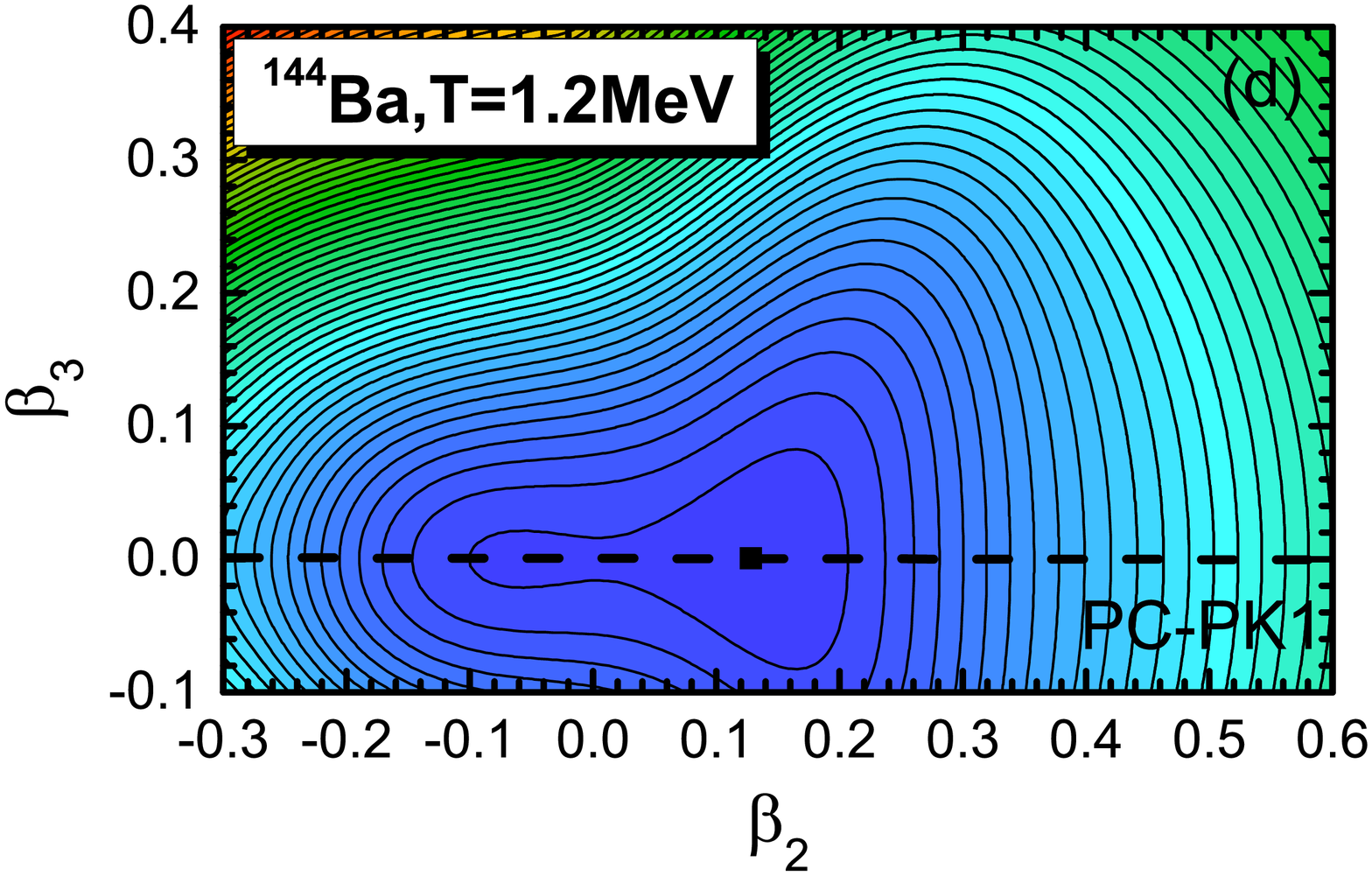} %
\includegraphics[scale=0.2]{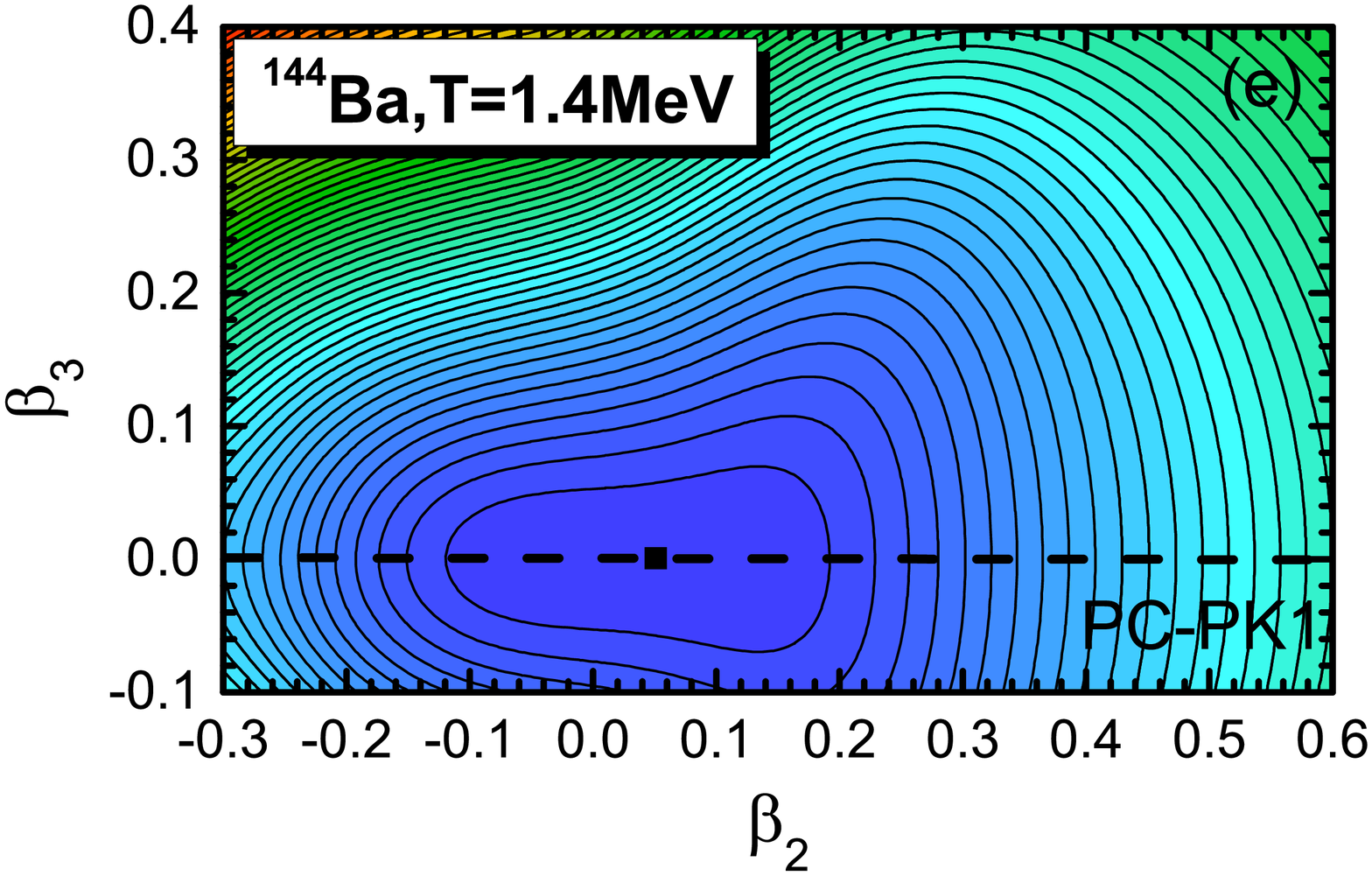} %
\includegraphics[scale=0.2]{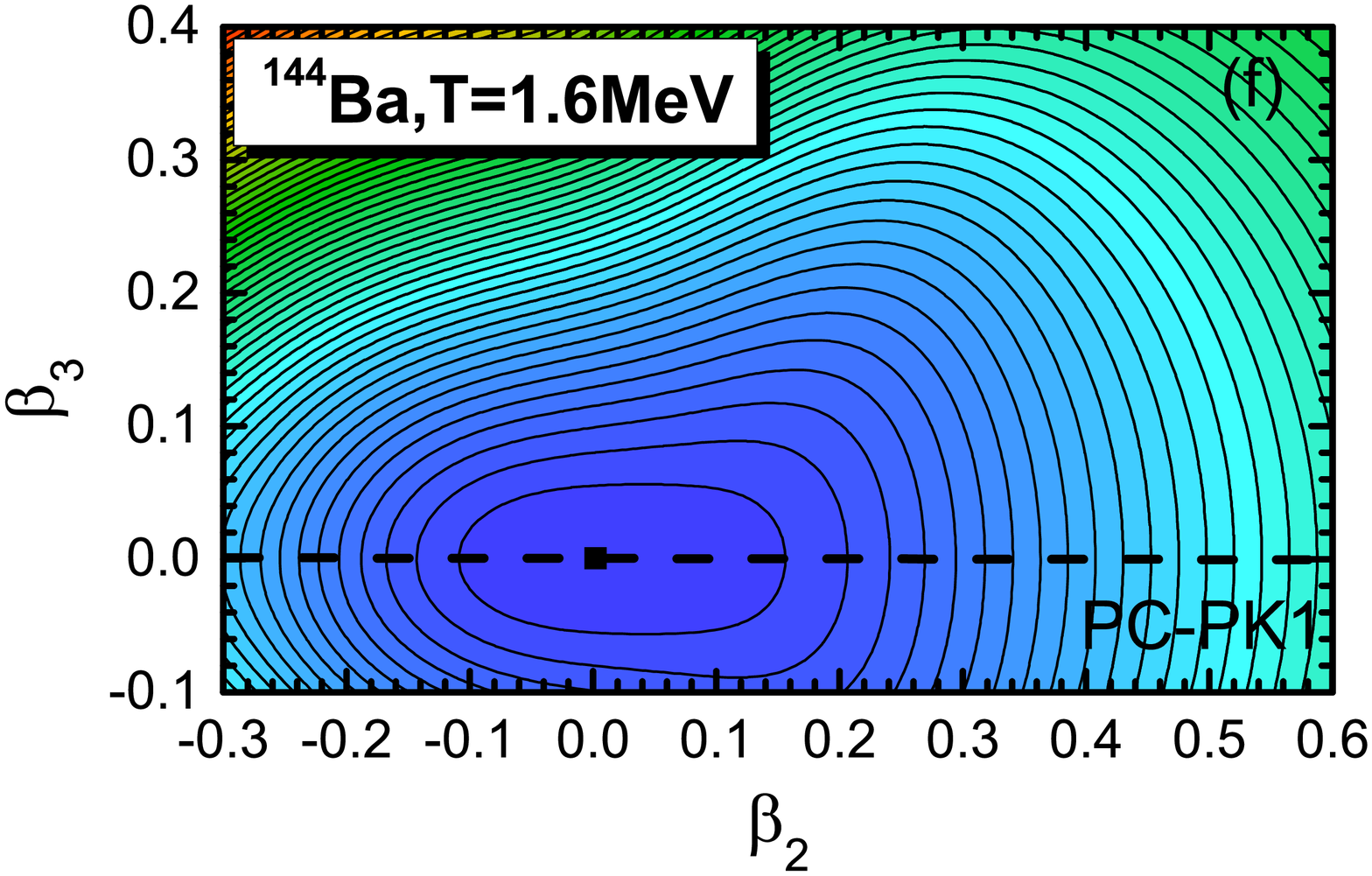} \\
\caption{(Color online) Same as Fig.~\ref{pes}, but
at temperatures $T$=0 (a), 0.4 (b), 0.8 (c), 1.2 (d), 1.4 (e), 1.6 (f) MeV for $^{144}$Ba.
}
\label{pes144}
\end{figure}
\begin{figure}[htb]
\includegraphics[scale=0.2]{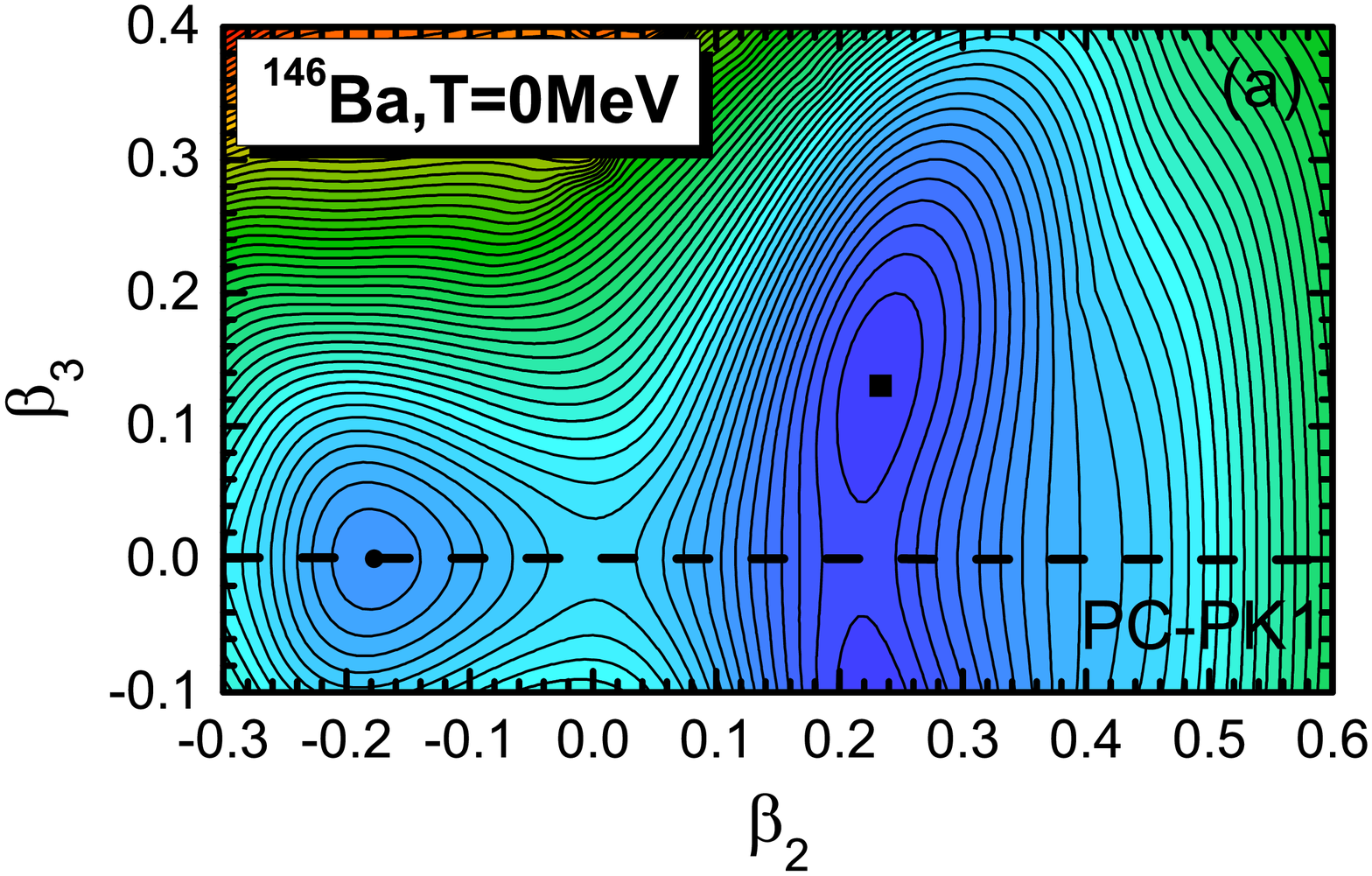} %
\includegraphics[scale=0.2]{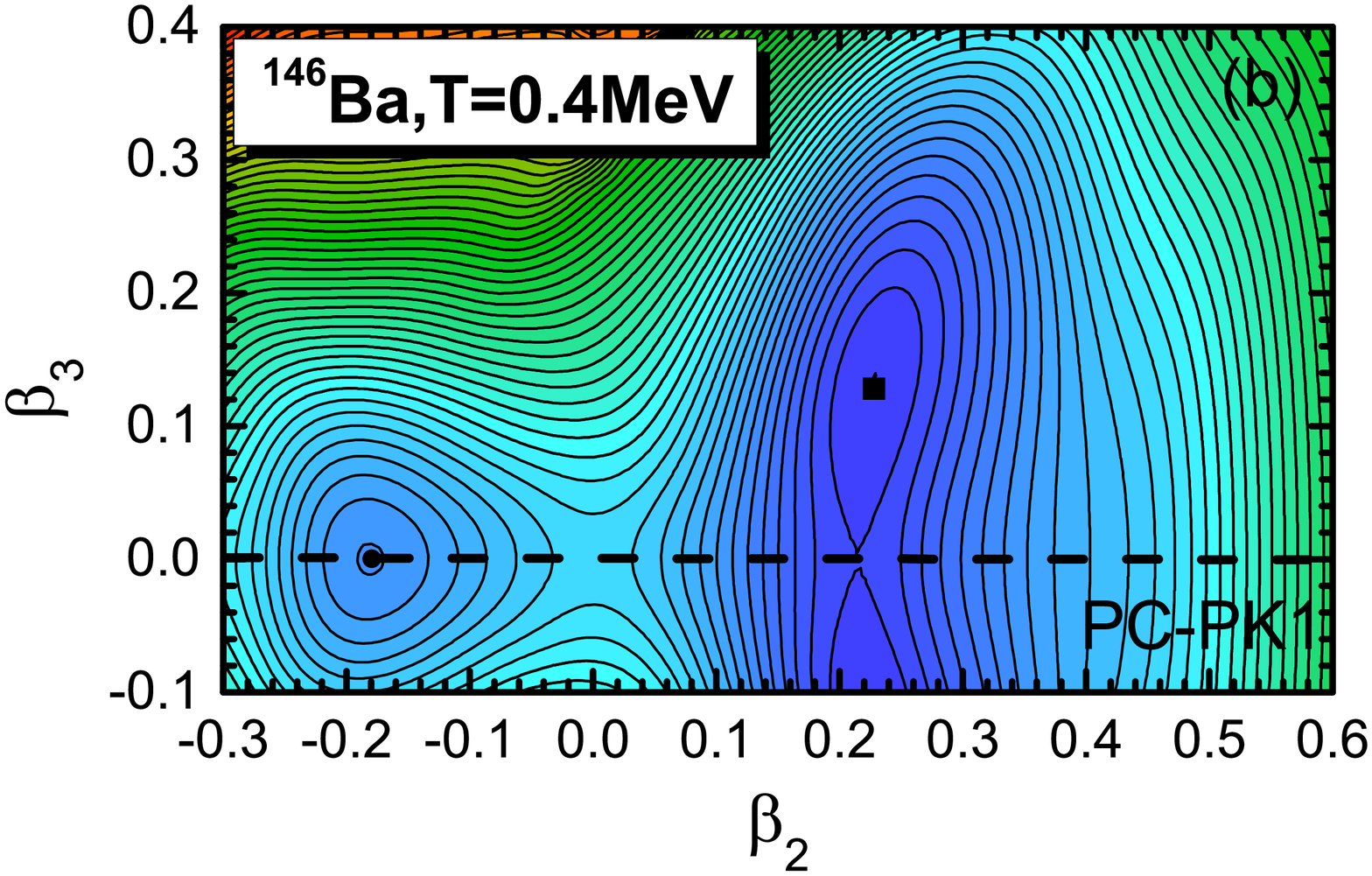} %
\includegraphics[scale=0.2]{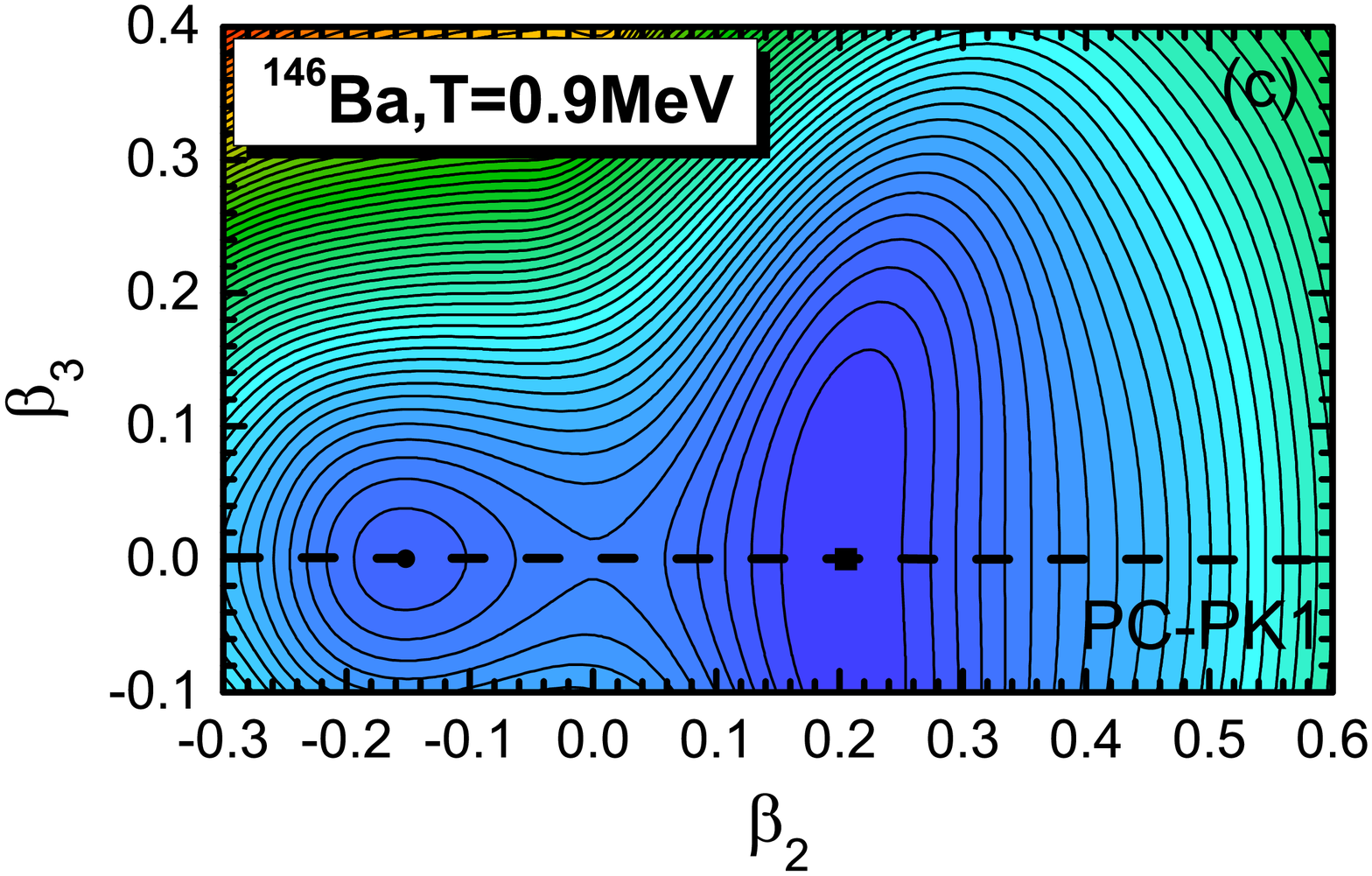} \\
\includegraphics[scale=0.2]{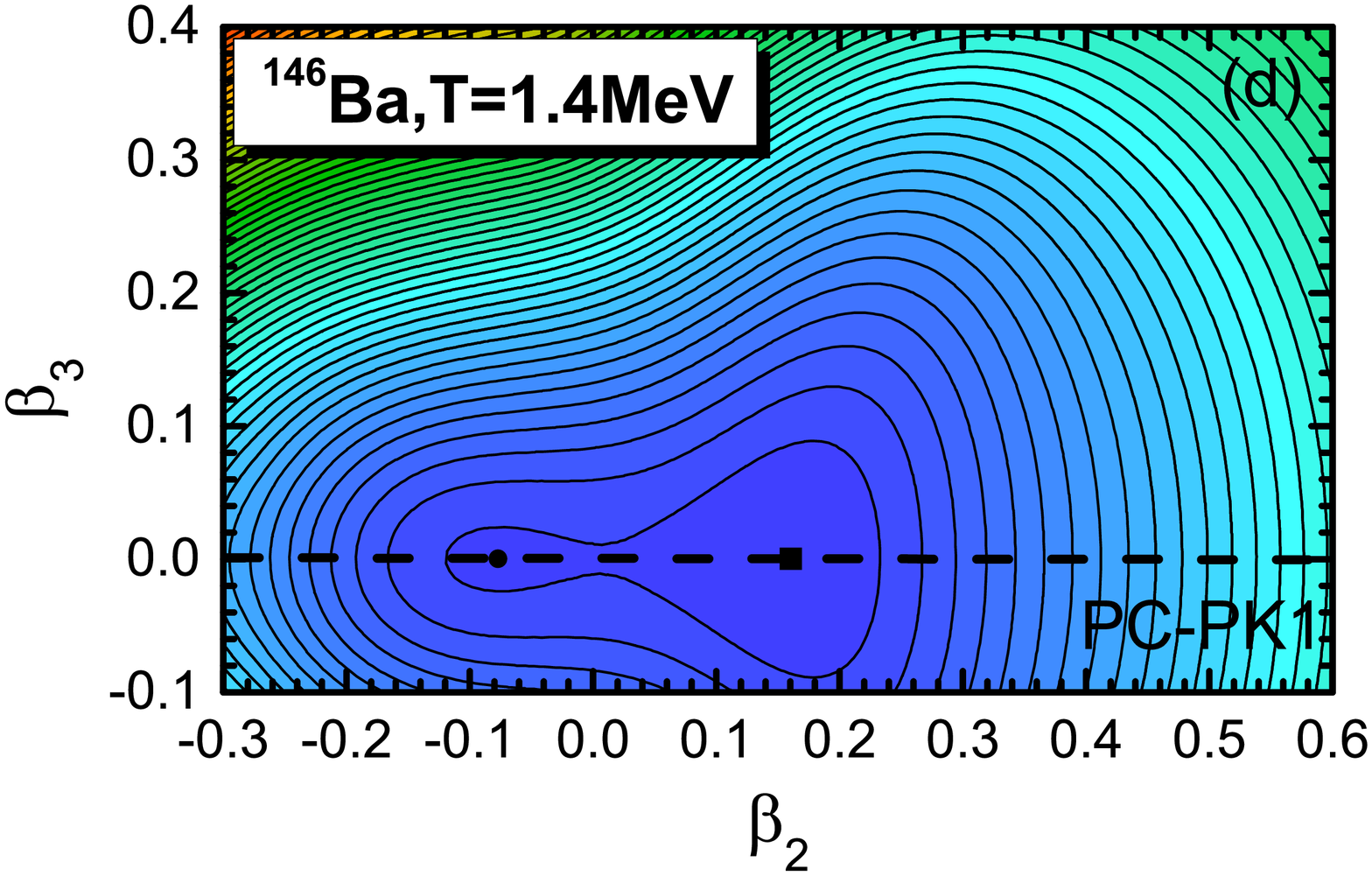} %
\includegraphics[scale=0.2]{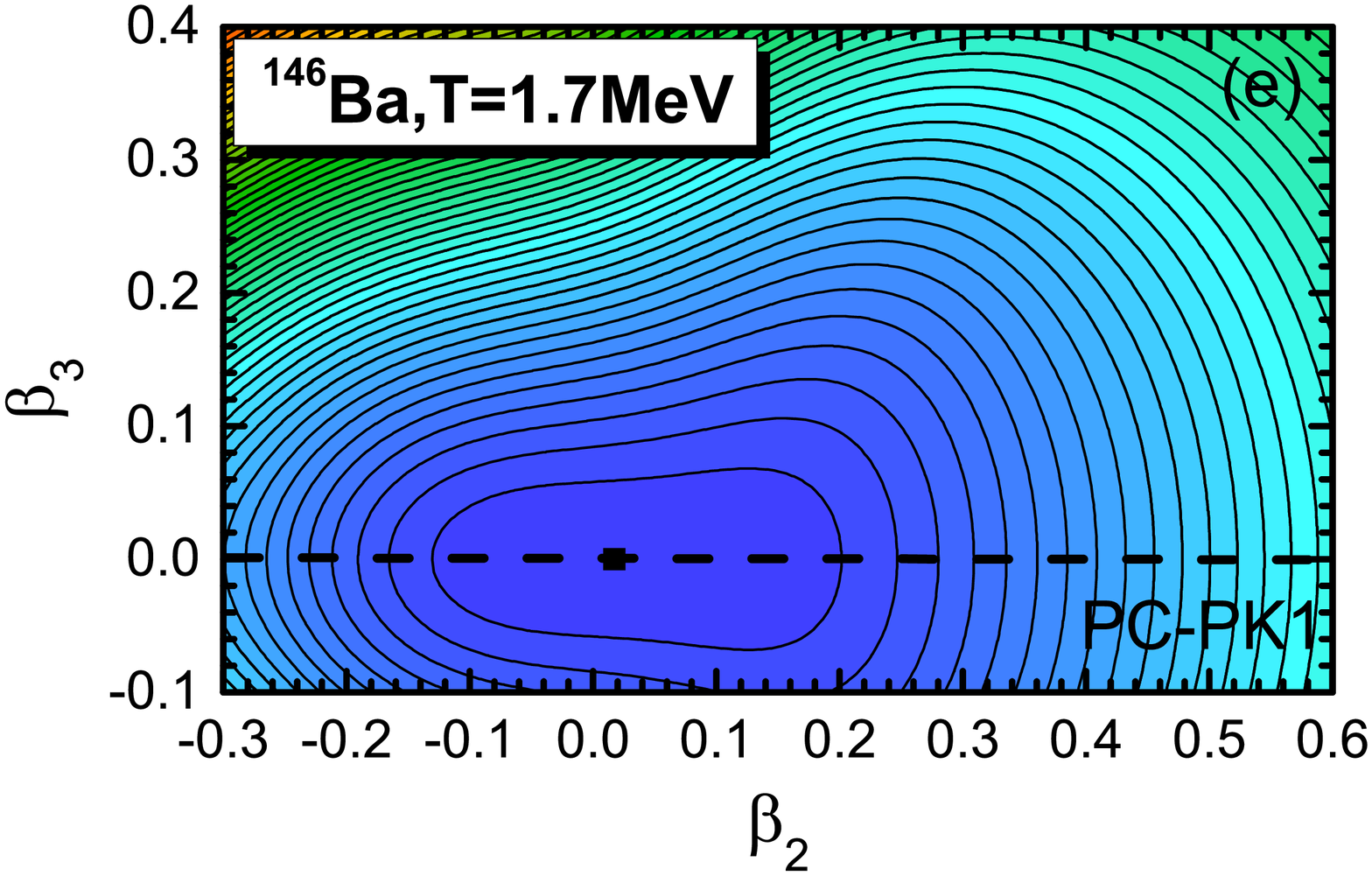} %
\includegraphics[scale=0.2]{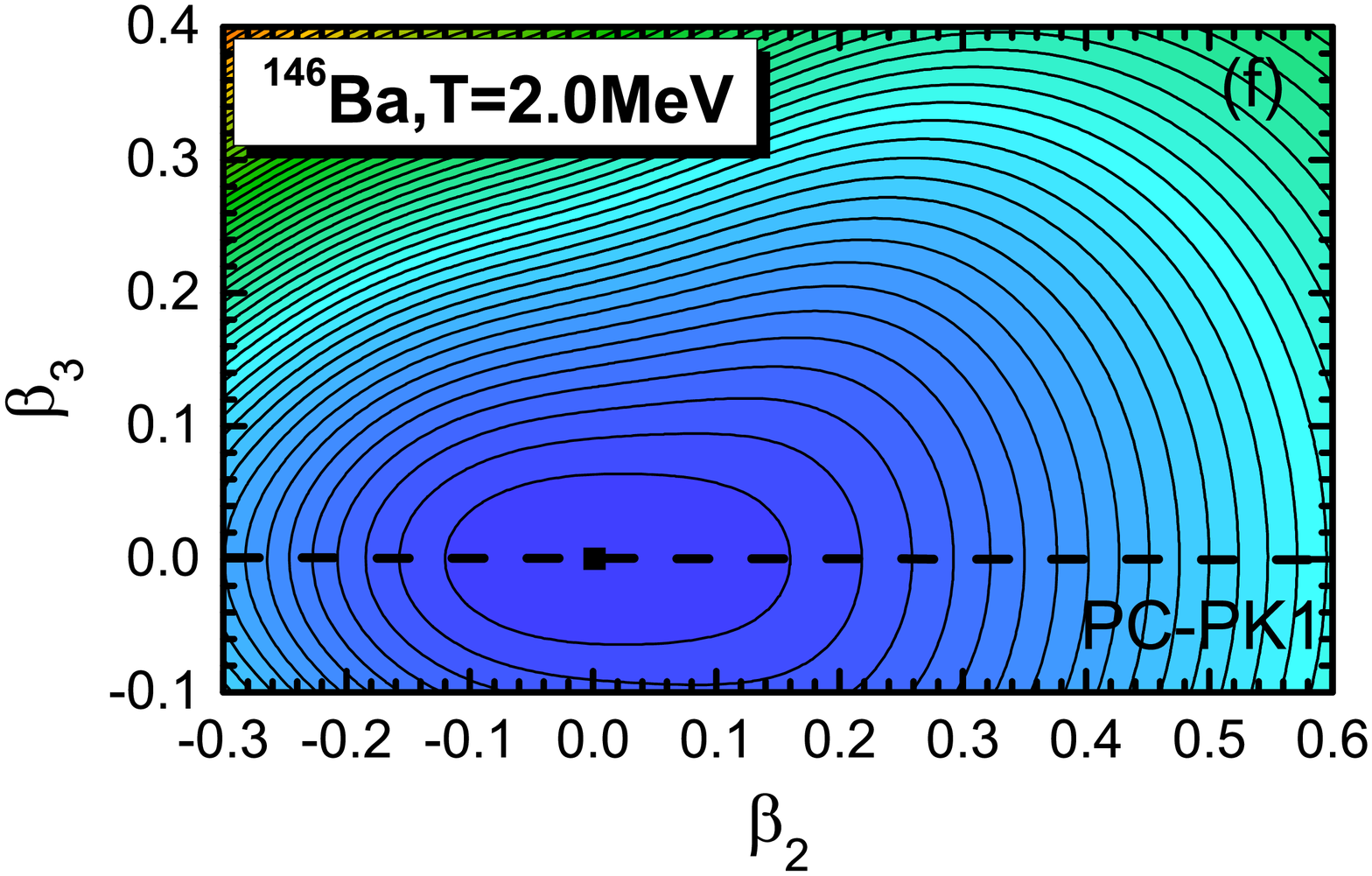} \\
\caption{(Color online) Same as Fig.~\ref{pes}, but
at temperatures $T$=0 (a), 0.4 (b), 0.9 (c), 1.4 (d), 1.7 (e), 2.0 (f) MeV for $^{146}$Ba.
}
\label{pes146}
\end{figure}

Additionally, we focus on neutron-rich $^{144}$Ba and $^{146}$Ba
where direct experimental evidence of octupole deformation was found recently~\cite{Bucher2016,Bucher2017}.
The free energies in the ($\beta_2$, $\beta_3$) plane at different temperatures
for $^{144,146}$Ba are plotted in Figs.~\ref{pes144} and \ref{pes146} respectively.
At zero temperature in Figs.~\ref{pes144}(a) and \ref{pes146}(a),
the octupole-deformed ground states of both $^{144}$Ba and $^{146}$Ba
are about 0.6 MeV lower than the quadrupole-deformed saddle points.
The calculated ground state quadrupole and octupole deformations are in good agreement with the experimental data.
For $^{144}$Ba, the experimental data are $\beta_2$=0.18, $\beta_3$=0.11-0.21~\cite{Bucher2016} while
the calculated ones are $\beta_2$=0.22, $\beta_3$=0.13.
With increasing temperature before the octupole transition temperature in Figs.~\ref{pes144}(b) and \ref{pes146}(b),
the energy differences between the global minimum and the saddle points become shallower.
At the octupole transition temperature in Figs.~\ref{pes144}(c) and \ref{pes146}(c),
the energy surface in the $\beta_3$ direction is softer than in the $\beta_2$ direction.
Note the oblate minima continuously become shallower in Figs.~\ref{pes144}(b),\ref{pes144}(c) and \ref{pes146}(b),\ref{pes146}(c).
At temperature higher than the octupole transition temperature in Figs.~\ref{pes144}(d) and \ref{pes146}(d),
a soft deformation area around the spherical state is developed with the disappearance of the oblate minima.
The transition from quadrupole deformed to spherical phase occurs in Figs.~\ref{pes144}(e) and \ref{pes146}(e),
while the spherical states are kept as the global minima in Figs.~\ref{pes144}(f) and \ref{pes146}(f).
It should be noted that the free energy surfaces near or at the transitions are
very similar for $^{144,146}$Ba, but the transition temperatures vary;
e.g., in Figs.~\ref{pes144}(e) and \ref{pes146}(e) the quadrupole transition temperatures for $^{144,146}$Ba
are 1.4 and 1.7 MeV respectively.

\begin{figure}[!htbp]
\includegraphics[scale=0.6]{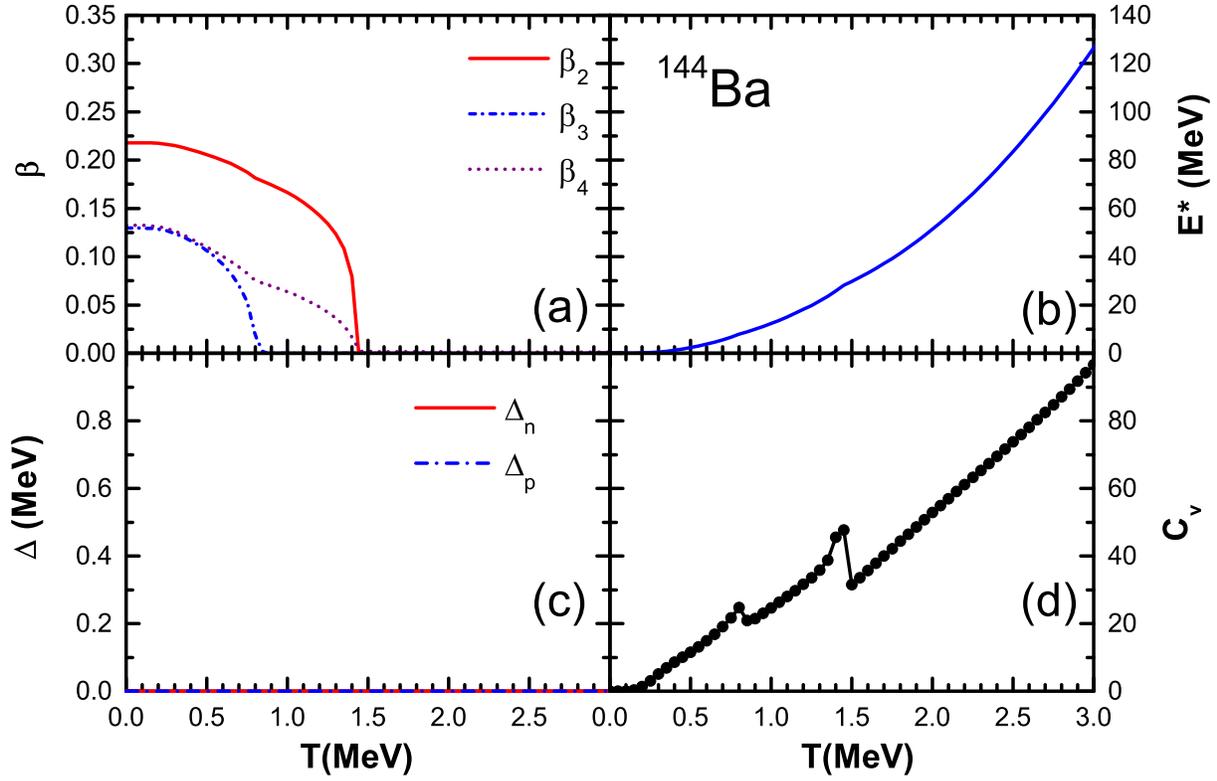}
\caption{(Color online) Same as Fig.~\ref{basic}, but for $^{144}$Ba.}
\label{Ba144}
\end{figure}
\begin{figure}[!htbp]
\includegraphics[scale=0.6]{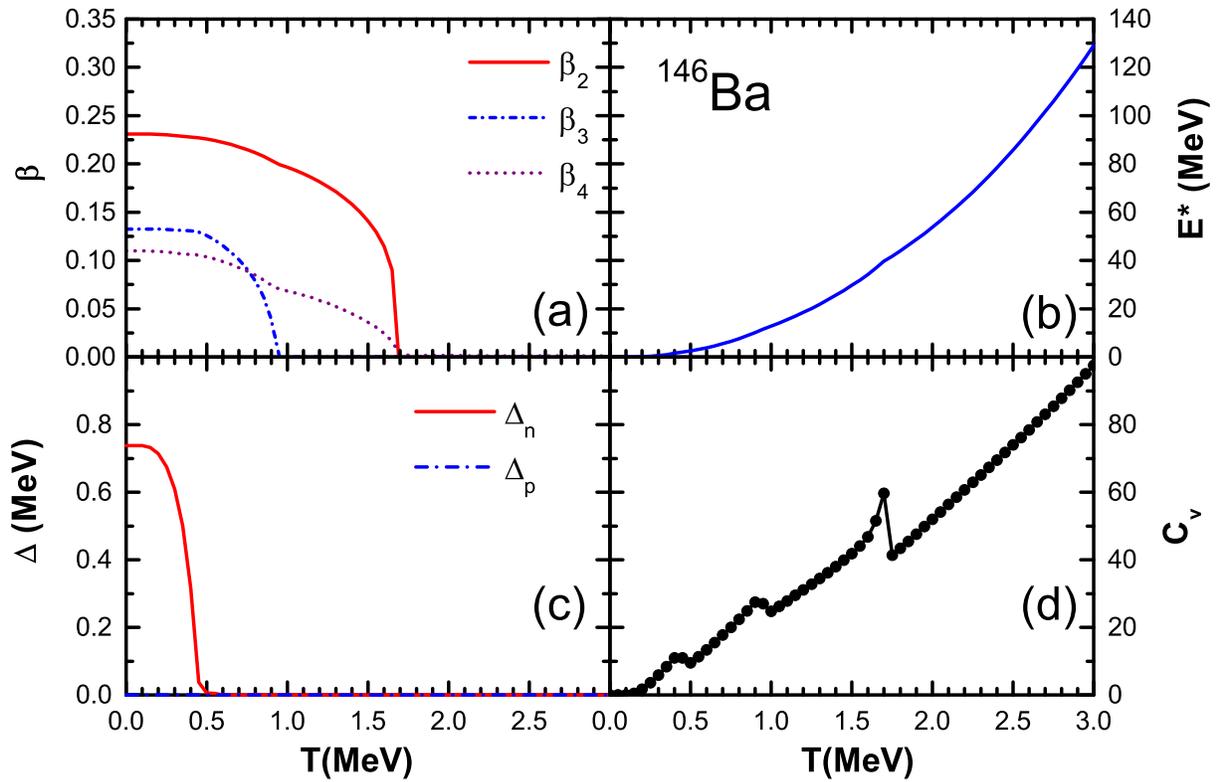}
\caption{(Color online) Same as Fig.~\ref{basic}, but for $^{146}$Ba.}
\label{Ba146}
\end{figure}
\begin{figure}[!htbp]
\includegraphics[scale=0.6]{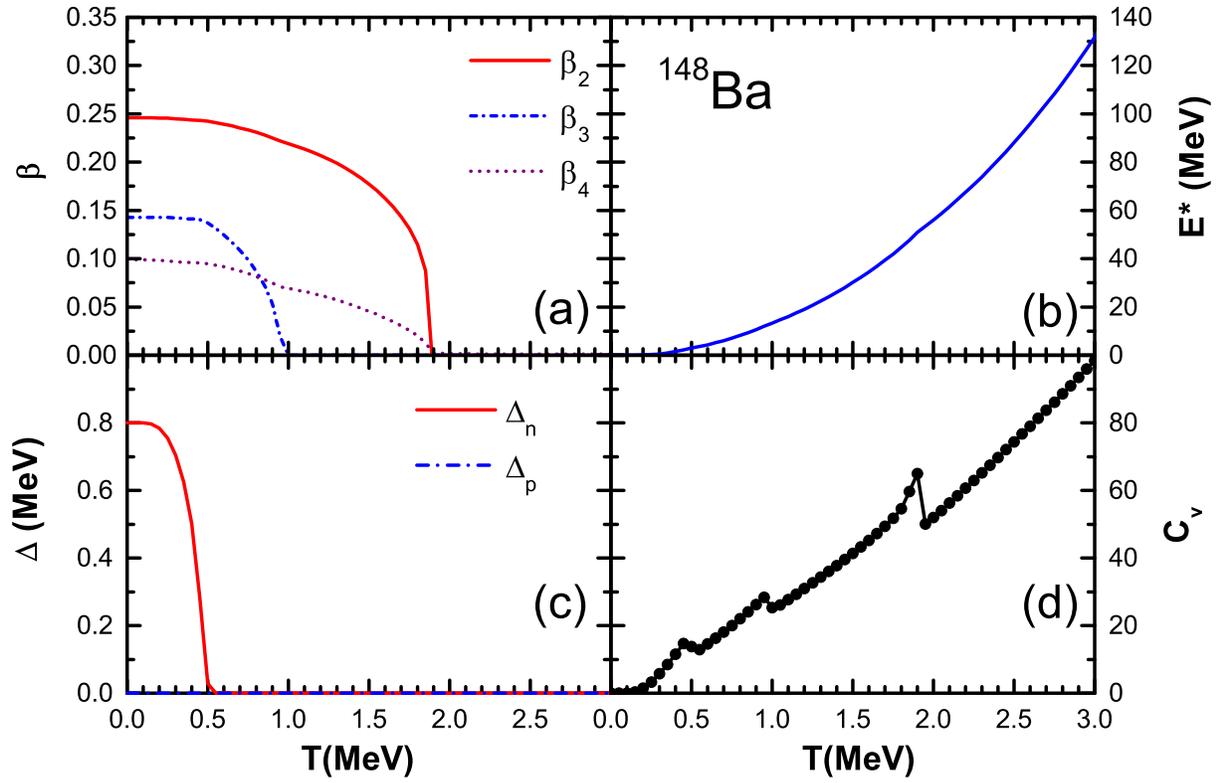}
\caption{(Color online) Same as Fig.~\ref{basic}, but for $^{148}$Ba.}
\label{Ba148}
\end{figure}
\begin{figure}[!htbp]
\includegraphics[scale=0.6]{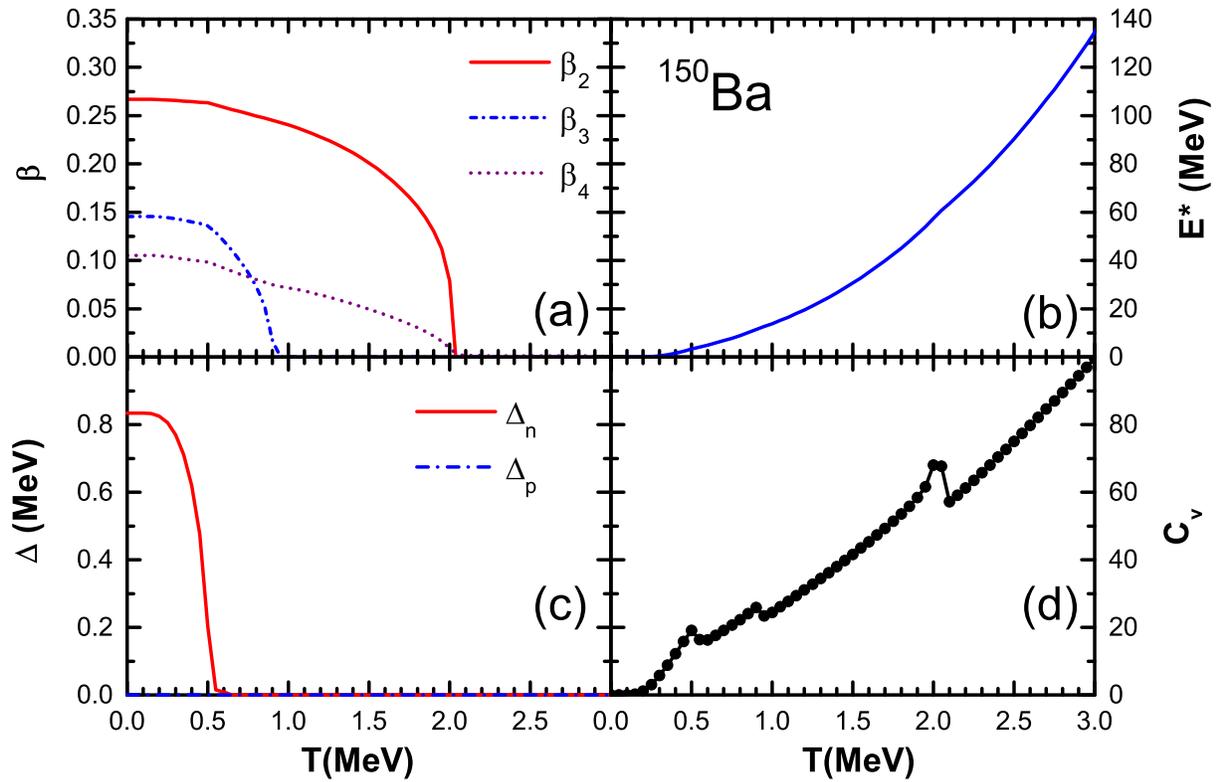}
\caption{(Color online) Same as Fig.~\ref{basic}, but for $^{150}$Ba.}
\label{Ba150}
\end{figure}
\begin{figure}[!htbp]
\includegraphics[scale=0.6]{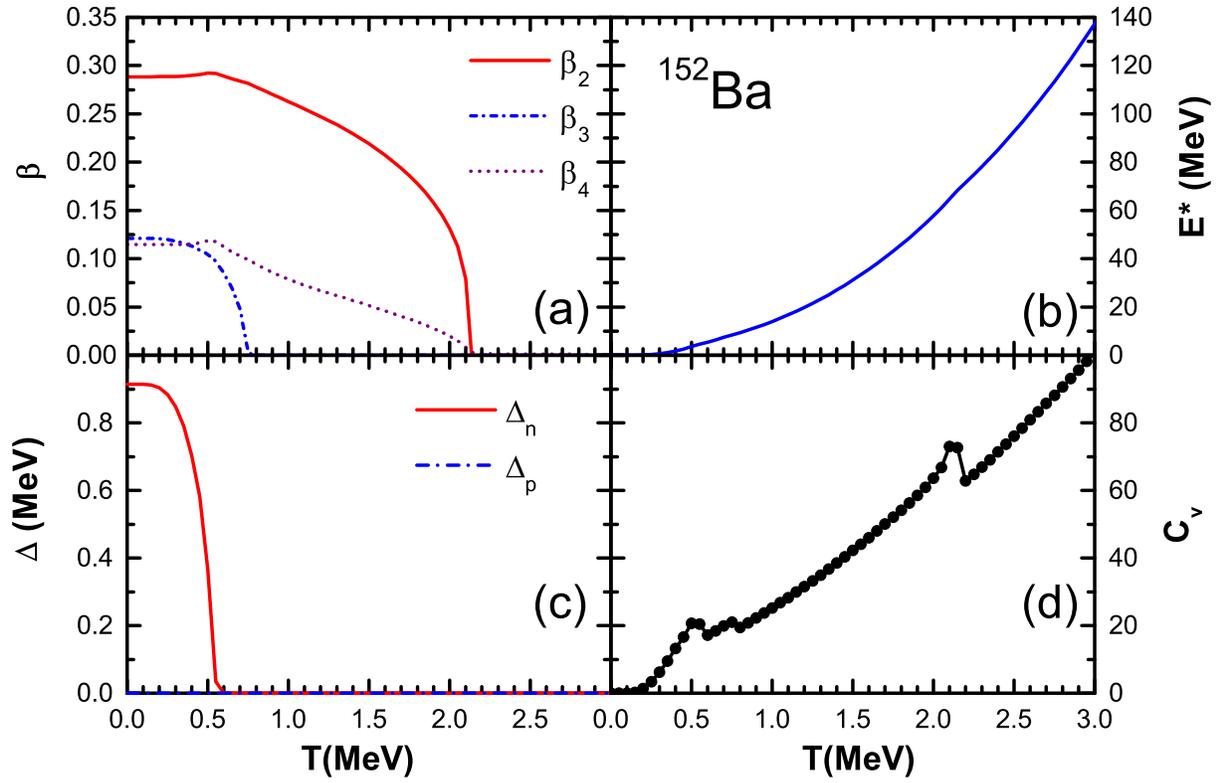}
\caption{(Color online) Same as Fig.~\ref{basic}, but for $^{152}$Ba.}
\label{Ba152}
\end{figure}
\begin{figure}[!htbp]
\includegraphics[scale=0.6]{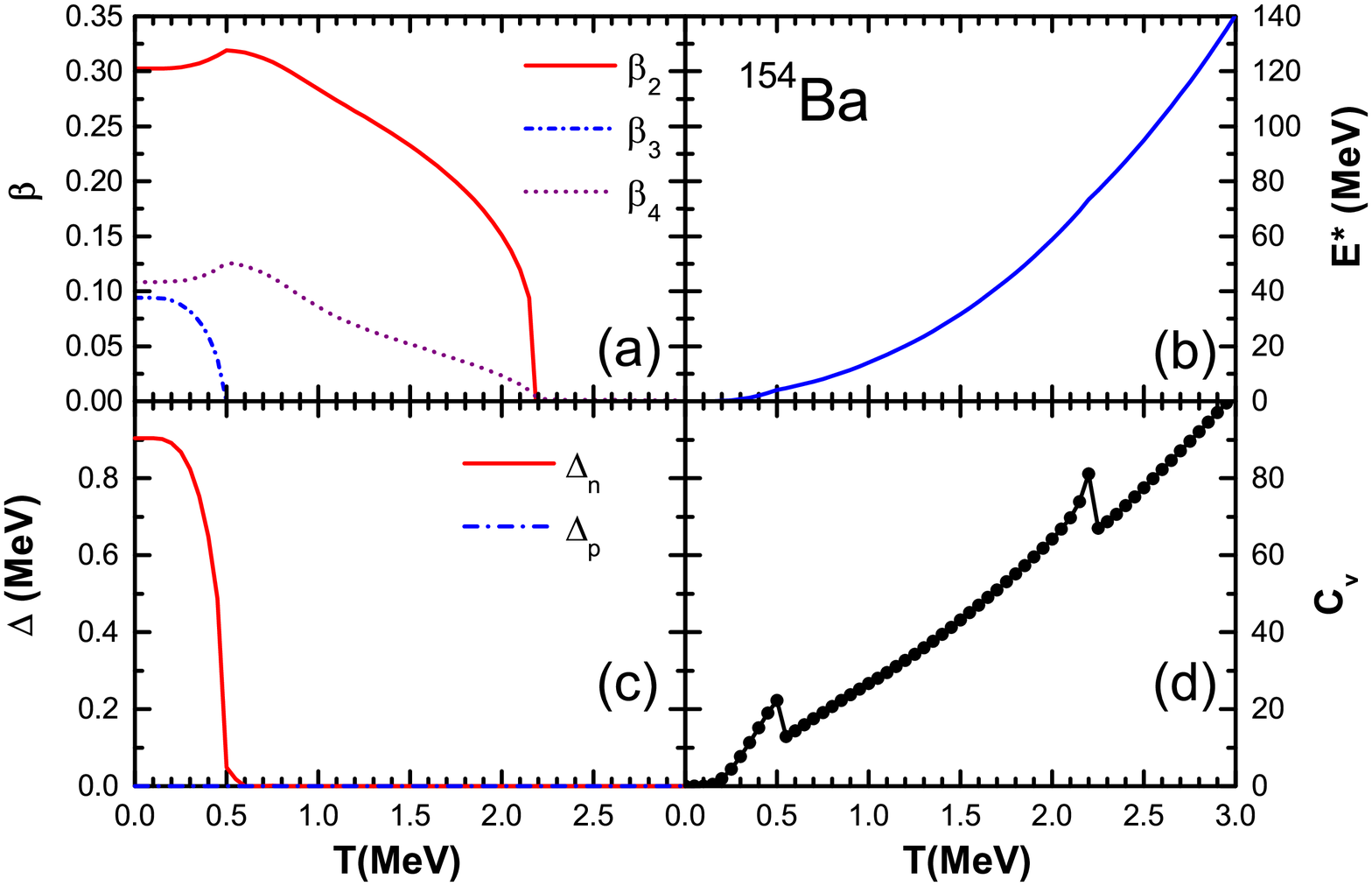}
\caption{(Color online) Same as Fig.~\ref{basic}, but for $^{154}$Ba.}
\label{Ba154}
\end{figure}

Moreover, even-even $^{144-154}$Ba isotopes are studied using the same method,
and their properties are shown in Figs.~\ref{Ba144}-\ref{Ba154}.
The quadrupole deformation for $^{144-154}$Ba ground states is increasing
since the neutron number is approaching the middle of a shell between the closures 82 and 126,
while the corresponding octupole deformation is decreasing
as the neutron number is departing from 88 where the strong octupole correlation is believed to occur~\cite{Butler1996}.
The results are also consistent with calculations of parameter sets PK1~\cite{Zhang2010}, DD-PC1~\cite{Ring2016}, and NL3*~\cite{Ring2016}.
For each isotope, when the temperature rises, similar to $^{224}$Ra, the global minima quadrupole and octupole deformations
evolve with the temperature similarly, where they change slowly for temperatures below the transition temperature
and then quickly drop to zero approaching the transition temperature.
Note the quadrupole and octupole transition temperatures for each isotope are separated.
The hexadecapole deformation drops little at the octupole transition temperature and vanishes at the quadrupole transition temperature.
Concerning the transition temperatures as functions of neutron number,
it is observed that the quadrupole transition temperature increases when the nuclei become heavier while
the octupole transition temperature remains nearly constant and finally decreases in Figs.~\ref{Ba144}(a) -\ref{Ba154}(a).
For the clarity, the quadrupole and octupole transition temperatures
together with the quadrupole and octupole deformations at the ground states are listed in Table~\ref{betaT}.
It is found that for even-even $^{144-154}$Ba isotopes,
the quadrupole and octupole transition temperatures are roughly linearly proportional to
the quadrupole and octupole deformations at the ground states.
Similar to the above discussion on $^{224}$Ra,
$^{144-154}$Ba may suffer from the neutron artificial shell $N$=92, getting lower transition temperatures.

\begin{table}\tabcolsep=4pt
\caption{
The quadrupole and octupole transition temperatures $T_2$ and $T_3$ (in MeV),
together with the quadrupole and octupole deformations $\beta_2$ and $\beta_3$ at the ground states
for even-even $^{144-154}$Ba,
obtained by constrained RMF+BCS calculations using the PC-PK1 energy density functional.}
\begin{tabular}{ccccc}
\hline\hline
Nucleus & $\beta_2$ & $T_2$(MeV) & $\beta_3$ & $T_3$(MeV) \\
\hline
$^{144}$Ba &    0.22  &  1.45  &  0.13  &   0.85 \\
$^{146}$Ba &    0.23  &  1.70  &  0.13  &   0.95 \\
$^{148}$Ba &    0.25  &  1.90  &  0.14  &   0.95 \\
$^{150}$Ba &    0.27  &  2.05  &  0.15  &   0.95 \\
$^{152}$Ba &    0.29  &  2.15  &  0.12  &   0.75 \\
$^{154}$Ba &    0.30  &  2.20  &  0.09  &   0.50 \\
\hline\hline
\label{betaT}
\end{tabular}
\end{table}

For the pairing correlations in Figs. ~\ref{Ba144}(c) -\ref{Ba154}(c),
only neutron pairing gaps of around 0.8 MeV survive except for pairing collapse in $^{144}$Ba.
All critical temperatures for pairing phase transitions follow the rule $T_c=0.6\Delta_n(0)$.
For the relative excitation energies in Figs.~\ref{Ba144}(b)-\ref{Ba154}(b),
some inappreciable kinks corresponding to the pairing and shape transitions can be found.
These kinks are manifested more clearly in the specific heat in Fig.~\ref{Ba144}(d)-\ref{Ba154}(d),
where three discontinuities corresponding to one pairing transition and two shape transition temperatures
are found except for $^{144,154}$Ba.
For $^{144}$Ba, the pairing transition is absent as a direct result of pairing collapse.
For $^{154}$Ba, the pairing transition coincides with the octupole shape transition.
The discontinuities of the specific heat in Figs.~\ref{Ba144}(d)-\ref{Ba154}(d) are consistent
with transition temperatures in Figs.~\ref{Ba144}(a)-\ref{Ba154}(a).

\section {Summary}
In summary,
the finite-temperature deformed RMF + BCS theory based on the relativistic point-coupling density functional
is applied to the shape evolution study of $^{224}$Ra and even-even $^{144-154}$Ba with temperature.
The free energy surfaces as well as the bulk properties including deformations,
pairing gaps, excitation energy, and specific heat for the global minimum are investigated.
For $^{224}$Ra, three discontinuities are found in the specific heat curve,
indicating the pairing transition at temperature 0.4 MeV,
and two shape transitions at temperatures 0.9 and 1.0 MeV, namely
one from quadrupole-octupole deformed to quadrupole deformed,
and the other from quadrupole deformed to spherical.
Furthermore, the single-particle levels as functions of the temperature are analyzed.
The gaps at $N=$136 and $Z=$88 are responsible for stabilizing the octupole-deformed global minimum at low temperatures.
With rising temperature, the shell effects disappear.
Similar pairing transition at $T\sim$0.5 MeV  and shape transitions at $T$=0.5$\sim$2.2 MeV
are found for even-even $^{144-154}$Ba.
Roughly there is a simple proportional relation between
the quadrupole and octupole transition temperatures and
the quadrupole and octupole deformations at the ground states.
The realistic description of statistical and quantal fluctuations is under considerations.


\section*{Acknowledgments}

We sincerely express our gratitude to Zhipan Li for helpful discussions.
This work was supported in part by
the National Natural Science Foundation of China under Grants No. 11105042, No. 11305161, and No. 11505157,
the China Scholarship Council,
and the Open Fund of Key Laboratory of Time and Frequency Primary Standards, Chinese Academy of Sciences.
The theoretical calculation was supported by the nuclear data storage system in Zhengzhou University.
Work at the Molecular Foundry, Lawrence Berkeley National Laboratory was supported by the Office of Science,
Office of Basic Energy Sciences, of the U.S. Department of Energy under Contract No. DE-AC02-05CH11231.


\begin{thebibliography}{90}

\vspace{3mm}

\bibitem{Egido1993}      J. L. Egido and P. Ring, J. Phys. G {\bf 19}, 1 (1993).
\bibitem{Egido2000}      J. L. Egido, L. M. Robledo, and V. Martin, Phys. Rev. Lett. {\bf 85}, 26 (2000).
\bibitem{GDR}            A. Schiller, and M. Thoennessen, At. Data Nucl. Data Tables {\bf 93}, 549 (2007).
\bibitem{Ring1984}       P. Ring, L. M. Robledo, J. L. Egido, and M. Faber, Nucl. Phys. {\bf A419}, 261 (1984).
\bibitem{Niu2009}        Y. F. Niu, N. Paar, D. Vretenar, and J. Meng, Phys. Lett. B {\bf 681}, 315 (2009).
\bibitem{Chak2016}       D. R. Chakrabarty, N. Dinh Dang, and V.M. Datar, Eur. Phys. J. A {\bf 52}, 143 (2016).

\bibitem{Bloch1958}     C. Bloch and C. Dedominicis, Nucl. Phys. {\bf 7}, 459 (1958). 
\bibitem{Sauer1976}     G. Sauer, H. Chandra, U. Mosel, Nucl. Phys. {\bf A264}, 221(1976). 
\bibitem{Lee1979}       H. C. Lee, and S. Das Gupta, Phys. Rev. C {\bf 19}, 2369 (1979).
\bibitem{Brack1974}     M. Brack and P. Quentin, Phys. Lett. {\bf 52B}, 159 (1974); Phys. Scr. {\bf A10}, 163 (1974). 
\bibitem{Quentin1978}   P. Quentin and H. Flocard, Annu. Rev. Nucl. Part. Sci. {\bf 28}, 523 (1978). 
\bibitem{Yen1994}       G. D. Yen, and H. G. Miller, Phys. Rev. C {\bf 50}, 807(1994).
\bibitem{Goodman1981}   A. L. Goodman, Nucl. Phys. {\bf A352}, 30 (1981).
\bibitem{Goodman1986}   A. L. Goodman, Phys. Rev. C {\bf 34}, 1942 (1986).
\bibitem{Egido2003}     V. Martin, J. L. Egido, and L. M. Robledo, Phys. Rev. C {\bf 68}, 034327 (2003).

\bibitem{Levit1984}      S. Levit and Y. Alhassid, Nucl. Phys. {\bf A413}, 439 (1984).
\bibitem{Alhassid1984}   Y. Alhassid and J. Zingman, Phys. Rev. C {\bf 30}, 684 (1984).
\bibitem{Rossignoli1994} R. Rossignoli and P. Ring, Ann. Phys. (N.Y.) {\bf 235}, 350 (1994).
\bibitem{Lang1993}       G. H. Lang, C.W. Johnson, S. E. Koonin, and W. E. Ormand, Phys. Rev. C {\bf 48}, 1518 (1993).
\bibitem{Dang1993}       N. D. Dang, P. Ring, and R. Rossignoli, Phys. Rev. C {\bf 47}, 606 (1993).
\bibitem{Dang2007}       N. D. Dang, Phys. Rev. C {\bf 76}, 064320 (2007).
\bibitem{Gambacurta2013} D. Gambacurta, D. Lacroix, and N. Sandulescu, Phys. Rev. C {\bf 88}, 88, 034324 (2013).
\bibitem{Liu2015}        L. Liu, Z. H. Zhang, and P. W. Zhao, Phys. Rev. C {\bf 92}, 044304 (2015).
\bibitem{Yuksel2014}     E. Y\"{u}ksel, E. Khan, K. Bozkurt, and G. Col\`{o}, Eur. Phys. J. A {\bf 50}, 160 (2014).


\bibitem{Bizzeti2003}    P. G. Bizzeti, in {\it Symmetries in Physics}, edited by A. Vitturi and R. Casten (World Scientific, Singapore, 2003), p. 262.
\bibitem{Bizzeti2005}    P. G. Bizzeti and A. M. Bizzeti-Sona, in {\it Nuclear Theory 24}, edited by S. Dimitrova (Heron Press, Sofia, 2005), p. 311.

\bibitem{Nature2013}    L. P. Gaffney {\it et al.}, Nature (London) {\bf 497}, 199 (2013).
\bibitem{Engel2013}     J. Engel, M. J. Ramsey-Musolf, and U. van Kolck, Prog. Part. Nucl. Phys. {\bf 71}, 21 (2013).

%

\bibitem{Ring1996}      P. Ring, Prog. Part. Nucl. Phys. {\bf 37}, 193 (1996).
\bibitem{Vretenar2005}  D. Vretenar, A. V. Afanasjev, G. A. Lalazissis, and P. Ring,Phys. Rep. {\bf 409}, 101 (2005).
\bibitem{Meng2006}      J. Meng, H. Toki, S.-G. Zhou, S. Q. Zhang, W. H. Long, and L. S. Geng, Prog. Part. Nucl. Phys. {\bf 57}, 470 (2006).

\bibitem{Niu2013}        Y. F. Niu, Z. M. Niu, N. Paar, D. Vretenar, G.H. Wang, J.S. Bai, and J. Meng, Phys. Rev. C {\bf 88}, 034308 (2013).
\bibitem{Long2015}       J. J. Li, J. Margueron, W. H. Long, and N. Van Giai, Phys. Rev. C {\bf 92}, 014302 (2015).
\bibitem{Agrawal2000}    B. K. Agrawal, Tapas Sil, J. N. De, and S. K. Samaddar, Phys. Rev. C {\bf 62}, 044307 (2000).
\bibitem{Agrawal2001}    B. K. Agrawal, Tapas Sil, S. K. Samaddar, and J. N. De, Phys. Rev. C {\bf 63}, 024002 (2001).
\bibitem{Bonche1984}     P. Bonche, S. Levit, and D. Vautherin, Nucl. Phys. {\bf A427}, 278 (1984); {\bf A436}, 265 (1985).
\bibitem{Lisboa2016}     R. Lisboa, M. Malheiro and B. V. Carlson, Phys. Rev. C {\bf 93}, 024321 (2016).

\bibitem{Kr2017}         W. Zhang and Y. F. Niu, Chin. Phys. C, {\bf 41}, 094102 (2017).
\bibitem{Bender2000}     M. Bender, K. Rutz, P.-G. Reinhard, and J. A. Maruhn, Eur. Phys. J. A {\bf 8}, 59 (2000).
\bibitem{Zhao2010}       P. W. Zhao, Z. P. Li, J. M. Yao, and J. Meng, Phys. Rev. C {\bf 82}, 054319 (2010).

\bibitem{Yao2015}        J. M. Yao, E. F. Zhou, and Z. P. Li, Phys. Rev. C {\bf 92}, 041304(R) (2015).
\bibitem{Ring2016}       S. E. Agbemava, A. V. Afanasjev, and P. Ring, Phys. Rev. C {\bf 93}, 044304 (2016).

\bibitem{Robledo2010}    L. M. Robledo, M. Baldo, P. Schuck, and X. Vi\~{n}as, Phys. Rev. C {\bf 81}, 034315 (2010).
\bibitem{Long2007}       W. H. Long, H. Sagawa, N. V. Giai, and J. Meng, Phys. Rev. C {\bf 76}, 034314 (2007).
\bibitem{Meng2016}       J. Meng, Relativistic Density Functional for Nuclear Structure (World Scientific, Singapore, 2016), p. 159.
\bibitem{Butler1996}     P. A. Butler and W. Nazarewicz, Rev. Mod. Phys. {\bf 68}, 349 (1996).
\bibitem{Bucher2016}     B. Bucher {\it et al.}, Phys. Rev. Lett. {\bf 116}, 112503 (2016).
\bibitem{Bucher2017}     B. Bucher {\it et al.}, Phys. Rev. Lett. {\bf 118}, 152504 (2017).
\bibitem{Zhang2010}      W. Zhang, Z. P. Li, and S. Q. Zhang, Chin. Phys. C {\bf 34}, 1094 (2010).




\end{thebibliography}
\end{document}